\documentclass[reprint,amsmath,amssymb,aps,floatfix,showkeys,showpacs,
]{revtex4-1}

\usepackage[T1]{fontenc}
\usepackage[utf8]{inputenc}

\usepackage{graphicx,colordvi,longtable}
\usepackage{dcolumn}
\usepackage{bm}

\begin{document}

\title{Energy invariant for shallow water waves and the Korteweg -- de Vries equation.\\ Is energy always an invariant?
}

\author{Anna Karczewska}
 \email{A.Karczewska@wmie.uz.zgora.pl}
\affiliation{Faculty of Mathematics, Computer Science and Econometrics\\ University of Zielona G\'ora, Szafrana 4a, 65-246 Zielona G\'ora, Poland}

\author{Piotr Rozmej}
 \email{P.Rozmej@if.uz.zgora.pl}
\affiliation{Institute of Physics, Faculty of Physics and Astronomy \\
University of Zielona G\'ora, Szafrana 4a, 65-246 Zielona G\'ora, Poland}
\date{\today} 

\author{Eryk Infeld} \email{Eryk.Infeld@ncbj.gov.pl}
\affiliation{National Centre for Nuclear Research, Hoża 69, 00-681 Warszawa, Poland}

\date{\today}

\begin{abstract}
It is well known that the KdV equation has an infinite set of conserved quantities.
The first three are often considered to represent mass, momentum and energy.
Here we try to answer the question of how this comes about,
and also how these KdV quantities relate to those of the Euler shallow water equations.
Here Luke's Lagrangian is helpful. We also consider higher order extensions of KdV.
Though in general not integrable, in some sense they are almost so, these with the accuracy of the expansion.
\end{abstract}



\maketitle

\section{Introduction} \label{intro}

There exists a vast number of papers dealing with the shallow water problem. Aspects of the propagation of weakly nonlinear, dispersive waves are still beeing studied.  Last year we published two articles  \cite{KRR,KRI} in which Korteveg--de Vries type equations were derived in weakly nonlinear, dispersive and long wavelength limit.
The second order KdV type equation was derived. 
The second order KdV equation \cite{MS,BS}, sometimes called "extended KdV equation", was obtained for the case with a flat bottom. 
In derivation of the new equation we adapted the method described in \cite{BS}. In  \cite{KRI}, an analytic solution of 
this equation in the form of a particular soliton was found, as well.

It is well known, see, e.g.\ \cite{Miura,MGK,DrJ,Newell85}, 
that for the KdV equation there exists an infinite number of invariants, that is, integrals over space of 
functions of the wave profile and its derivatives, which are constants in time. Looking for analogous invariants for the second order KdV equation we met with some problems even for the standard KdV equation (which is first order in small parameters). This problem appears when  energy conservation is considered. 

In this paper we reconsider invariants of the KdV equation and formulas for the total energy
in several different approaches and different frames of reference (fixed and moving ones).
We find that the invariant ~$I^{(3)}$, sometimes called the energy invariant, does not always have that interpretation.  
We also give a proof that for the second order KdV equation, obtained in \cite{KRR,KRI,MS,BS},
~$\int_{-\infty}^{\infty} \eta^2 dx$ is not an invariant of motion.

There are many papers considering higher-order KdV type equations. Among them we would like to point out works of Byatt-Smith \cite{B-S87}, Kichenassamy and Olver \cite{SK_PO}, Marchant \cite{MS,MS96,Mar99,Mar02,Mar02a}, Zou and Su \cite{ZouSu},  Tzirtzilakis {\it et.al.}   \cite{TMAB} and Burde \cite{Burde}. It was shown that if some coefficients of the second order equation for shallow water problem (\ref{etaab}) are diferent or zero then there exists a hierarchy of solition solutions.  Kichenassamy and Olver \cite{SK_PO} even claimed that for second order KdV equation solitary solutions of appropriate form can not exist. This claim was falsified in our paper \cite{KRI} where the analytic solution of the second order KdV equation (\ref{etaab}) was found. Concerning the energy conservation there are indications that collisions of solitons \cite{Hirota72,Hir} which are solutions of higher order equations of KdV type can be inelastic \cite{ZouSu,TMAB}.

The paper is organized as follows. In Section \ref{equations} several frequently used forms of KdV equations are recalled with particular attention to transformations between fixed and moving reference frames. In Section \ref{invariants} the  form of the three lowest invariants of KdV equations is derived for different forms of the equations. In Section 
\ref{energy} we show that the energy calculated from the definition $H=T+V$ has no invariant form. Section \ref{varia} describes the variational approach in a potential formulation which gives a proper KdV equation but fails in obtaining second order KdV equations. In the next section the proper invariants are obtained from Luke's Lagrangian density. 
Section  \ref{concl} summarizes conclusions on the energy for KdV equation. In section \ref{extended} we apply the same formalism to calculate energy for waves governed by the extended KdV equation (second order). We found that energy is not conserved neither in fixed coordinate system nor in the moving frame. 


\section{The extended KdV equation} \label{equations}

\begin{figure}[tbh]
\begin{center}
\resizebox{1.0\columnwidth}{!}{\includegraphics{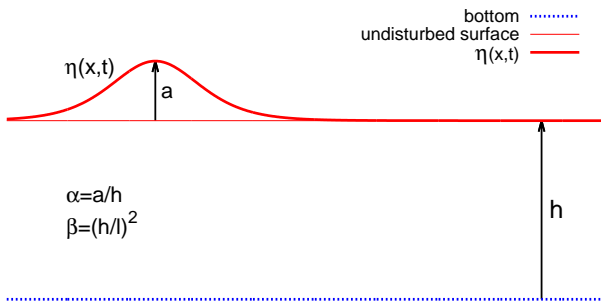}}
\end{center}
\vspace{-5mm}
\caption{Schematic view of the geometry.} 
 \label{geom}
\end{figure}

The geometry of shallow water waves is presented in Fig.~\ref{geom}.

In  \cite{KRR,KRI} we derived an equation, second order in small parameters, in the fixed reference system and with scaled nondimensional variables containing terms for bottom fluctuations. They will not be considered here.

For a flat bottom
 that equation reduces to the second order KdV type equation, identical with~\cite[Eq.~(21)]{BS} for $\beta=\alpha$, that is,
\begin{eqnarray} \label{etaab}
\eta_t+\eta_x + \alpha\, \frac{3}{2}\eta\eta_x +\beta\,\frac{1}{6} \eta_{3x} +
\alpha^2\,\left(-\frac{3}{8}\eta^2\eta_x\right) &&\\
 + \alpha\beta\,\left(\frac{23}{24}\eta_x\eta_{2x}+\frac{5}{12}\eta\eta_{3x} \right)+\beta^2\,\frac{19}{360}\eta_{5x} &=&0.  \nonumber
\end{eqnarray}
Subscripts denote partial differentiation.
Small parameters ~$\alpha,\beta$ 
are defined by ratios of the wave amplitude
 ~$a$, the average water depth ~$h$ and mean wavelength ~$l$ 
$$ \alpha =\frac{a}{h}, \qquad \beta=\left(\frac{h}{l}\right)^2. $$

Equation (\ref{etaab}) was earlier derived in \cite{MS} and called "the extended KdV equation".

Limitation to the first order in small parameters yields the KdV equation in a fixed coordinate system
\begin{equation} \label{kdv1}
\eta_t+\eta_x + \alpha\, \frac{3}{2}\eta\eta_x +\beta\,\frac{1}{6} \eta_{3x} =0.
\end{equation}
Transformation to a {\sl moving frame}~ in the form
\begin{equation} \label{tr1}
\bar{x} =(x-t), \qquad \bar{t}=t, \qquad \bar{\eta} = \eta,
\end{equation}
allows us to remove the term ~$\eta_{x}$ in  the KdV equation in a frame moving withthe velocity of sound $\sqrt{gh}$
\begin{equation} \label{kdv1m}
\bar{\eta}_{\bar{t}}+ \alpha\, \frac{3}{2}\bar{\eta}\bar{\eta}_{\bar{x}} +\beta\,\frac{1}{6} \bar{\eta}_{3\bar{x}} =0.
\end{equation}
The explicit form of the scaling leading to equations (\ref{etaab}) -- (\ref{kdv1m}) is given by (\ref{Przeskalowanie}).

Problems with mass, momentum and energy conservation in the KdV equation were  discussed in \cite{Kalisch} recently. In this paper the authors considered the KdV equations in the original dimensional variables. Then the KdV equatios are  
\begin{equation} \label{nieruch} 
\eta_t + c\eta_x +\frac{3}{2}\frac{c}{h}\eta\eta_x +\frac{c h^2}{6}\eta_{xxx}=0,
\end{equation}
in a fixed frame of reference and
\begin{equation} \label{ruch} 
\eta_t +\frac{3}{2}\frac{c}{h}\eta\eta_x +\frac{c h^2}{6}\eta_{xxx}=0,
\end{equation}
in a moving frame. In both, $c=\sqrt{gh}$, and (\ref{ruch}) is obtained from (\ref{nieruch}) by setting $x'=x-ct$ and dropping the prime sign.

In our present paper we discuss the energy formulas obtained both in fixed and moving frames of reference for KdV  (\ref{kdv1}), (\ref{kdv1m}), (\ref{nieruch}), (\ref{ruch}) . There seem to be some contradictions in the literature because the form of some invariants and the energy formulas differ in different sources  because of using different reference frames and/or different scalings. 
In this paper we address this problems.

The second goal is to 
 present some invariants for a KdV type equation of the second order (\ref{etaab}).

\section{Invariants of KdV type equations} \label{invariants}
What invariants can be attributed to equations  (\ref{etaab}) -- (\ref{kdv1}) and (\ref{nieruch}) -- (\ref{ruch}) ?

It is well known, see, e.g. \cite[Ch.~5]{DrJ}, that an equation 
of the form
\begin{equation} \label{coneq}
\frac{\partial T}{\partial t} + \frac{\partial X}{\partial x} =0,  
\end{equation}
where neither ~$T$ (an analog to density) nor  ~$X$ (an analog to flux) contain partial derivatives with respect to ~$t$, corresponds to some {\sl conservation law}. It can be applied, in particular, to KdV equations (where there exist an infinite number of such conservation laws) and to the equations of KdV type  like (\ref{etaab}). Functions ~$T$ and ~$X$ may depend on ~$x,t,\eta,\eta_x,\eta_{2x},\ldots,h, h_x,\ldots ,$ but not $\eta_t$. If both functions ~$T$ and ~$X_x$ are integrable on ~$(-\infty,\infty)$ and $\displaystyle\lim_{x\to\pm\infty} X=$ const (soliton solutions), then integration of equation (\ref{coneq}) yields
\begin{equation} \label{coneq1}
\frac{\mathrm{d}}{\mathrm{d} t}\left(\int_{-\infty}^{\infty} T\,dx \right)=0
\qquad \mbox{or} \qquad   \int_{-\infty}^{\infty} T\,dx = \mbox{const. } ,
\end{equation}
since
\begin{equation} \label{limits}
 \int_{-\infty}^{\infty} X_x\,dx = X(\infty,t)- X(-\infty,t) =0.
\end{equation}
The same conclusion applies for periodic solutions (cnoidal waves), when in the integrals  (\ref{coneq1}),  (\ref{limits}) limits of integration $(-\infty,\infty)$ are replaced by $(a,b)$, where $b-a=\Lambda$ is the space period of the cnoidal wave (the wave length).

\subsection{Invariants of the KdV equation} \label{niezm1}

For the KdV equation (\ref{kdv1}) the two first invariants can be obtained easily.
Writing  (\ref{kdv1}) in the form
\begin{equation} \label{e1}
\frac{\partial \eta}{\partial t} + \frac{\partial }{\partial x}\left(\eta+\frac{3}{4} \alpha \eta^2 +\frac{1}{6} \beta  \eta_{xx} \right) =0.
\end{equation}
one immediately obtains the conservation of mass (volume) law
\begin{equation} \label{I1}
I^{(1)} = \int_{-\infty}^{\infty} \eta\, dx = \mbox{const}. 
\end{equation}
Similarly, multiplication of (\ref{kdv1}) by ~$\eta$~ gives
\begin{equation} \label{e2}
 \frac{\partial }{\partial t} \left(\frac{1}{2} \eta^2 \right)+ \frac{\partial }{\partial x}\left(\frac{1}{2}\eta^2+\frac{1}{2} \alpha \eta^3 - \frac{1}{12} \beta \eta_x^2
 +\frac{1}{6} \beta \eta  \eta_{xx} \right) =0,
\end{equation}
resulting in the invariant of the form
\begin{equation} \label{I2}
 I^{(2)} = \int_{-\infty}^{\infty} \eta^2\, dx = \mbox{const}. 
\end{equation}
In the literature of the subject, see, e.g.\  \cite{Kalisch,DrJ}, ~$ I^{(2)}$ is attributed to momentum conservation.

Invariants $I^{(1)},I^{(2)}$ have the same form for all KdV equations (\ref{kdv1}), (\ref{kdv1m}), (\ref{kdvm}), (\ref{nieruch}), (\ref{ruch}).

Denote the left hand side of  (\ref{kdv1}) by ~$\mbox{KDV}(x,t)$ and take
\begin{equation} \label{cstI3}
3 \eta^2 \times \mbox{KDV}(x,t) -\frac{2}{3}\frac{\beta}{\alpha} \eta_x \times \frac{\partial} {\partial x} \mbox{KDV}(x,t).
\end{equation}
The result, after simplifications is 
\begin{eqnarray} \label{i3a}
\frac{\partial}{\partial t}\left( \eta^3 -\frac{1}{3} \frac{\beta}{\alpha}  \eta_x^2 \right)+ 
\frac{\partial}{\partial x}\left(
\frac{9}{8} \alpha \eta^4
+\frac{1}{2} \beta  \eta_{2x} \eta^2  \right. \hspace{7ex} && \\ \left.
- \beta \eta_x^2 \eta  +  \eta^3 
+\frac{1}{18} \frac{\beta^2}{\alpha} \eta_{2x}^2
-\frac{1}{9} \frac{\beta ^2}{\alpha} \eta_x \eta_{3x} 
-\frac{1}{3}  \frac{\beta}{\alpha}  \eta_x^2 \right) &=&0. \nonumber
\end{eqnarray}
Then the next invariant for KdV  in the fixed reference frame  (\ref{kdv1}) is
\begin{equation} \label{i3E}
 I^{(3)}_{\mathrm{fixed~frame}} = \int_{-\infty}^{\infty} \left(\eta^3 -\frac{1}{3}  \frac{\beta}{\alpha} \eta_x^2 \right) dx = \mbox{const}. 
\end{equation}

The same invariant is obtained for the KdV in the moving frame (\ref{kdv1m}). The same construction like (\ref{cstI3}) but for equation (\ref{kdv1m}) 
results in
\begin{eqnarray} \label{i3am}
\frac{\partial}{\partial t}\left( \eta^3 -\frac{1}{3} \frac{\beta}{\alpha}  \eta_x^2 \right)+
\frac{\partial}{\partial x}\left(
\frac{9}{8} \alpha \eta^4
+\frac{1}{2} \beta  \eta_{2x} \eta^2 \hspace{7ex} \right. &&\\ \left.
- \beta \eta_x^2 \eta 
+  \eta^3
+\frac{1}{18} \frac{\beta^2}{\alpha} \eta_{2x}^2
-\frac{1}{9} \frac{\beta ^2}{\alpha} \eta_x \eta_{3x} \right) &=&0. \nonumber
\end{eqnarray}
Then the next invariant for KdV equation in moving reference frame  (\ref{kdv1}) is
\begin{equation} \label{i3Em}
 I^{(3)}_{\mathrm{moving~frame}} = \int_{-\infty}^{\infty} \left(\eta^3 -\frac{1}{3}  \frac{\beta}{\alpha} \eta_x^2 \right) dx = \mbox{const}. 
\end{equation}

The procedure similar to those described above leads to the same invariants for both equations (\ref{nieruch}) and (\ref{ruch}) where KdV equations are written in dimensional variables. To see this, it is enough to take
 ~$3\eta^2\times kdv(x,t) -\frac{2}{3}h^3 \frac{\partial} {\partial x}  kdv(x,t)=0 $, where ~$kdv(x,t)$ is the lhs either of (\ref{nieruch}) or (\ref{ruch}).
For equation (\ref{nieruch}) the conservation law is
\begin{eqnarray} \label{nieruch1} 
\frac{\partial} {\partial t}\left(\eta^3-\frac{h^3}{3}\eta_x^2 \right)+ \frac{\partial} {\partial x}\left( c\eta^3-\frac{9c}{8h}\eta^4-\frac{1}{3}ch^3\eta_x^2 \right. \hspace{5ex} &&\\ \left.
-ch^2\eta\eta_x^2 +\frac{1}{2}ch^2\eta^2\eta_{xx}+\frac{1}{18}ch^5\eta_{xx}^2-\frac{1}{9}ch^5\eta_x\eta_{xxx}
\right)  &=&0, \nonumber
\end{eqnarray}
whereas for  equation (\ref{ruch}) the flux term is different
\begin{eqnarray} \label{ruch1} 
\frac{\partial} {\partial t}\left(\eta^3-\frac{h^3}{3}\eta_x^2 \right)+ \frac{\partial} {\partial x}\left( \frac{9c}{8h}\eta^4-ch^2\eta\eta_x^2\right.  \hspace{6ex} &&\\ \left.
+\frac{1}{2}ch^2\eta^2\eta_{xx}
+\frac{1}{18}ch^5\eta_{xx}^2-\frac{1}{9}ch^5\eta_x\eta_{xxx}
\right) &=&0. \nonumber
\end{eqnarray}
But in both cases the same $I^{(3)}$  invariant is obtained as
\begin{equation} \label{I3Kal} 
I^{(3)}_\mathrm{dimensional} =\int_{-\infty}^{\infty} \left(\eta^3-\frac{h^3}{3}\eta_x^2 \right) dx = \mbox{const}.
\end{equation}

\noindent
{\bf Conclusion~} {\sl Invariants $I^{(3)}$  have the same form for fixed and moving frames of reference when the transformation from fixed to moving frame scales ~$x$ and $t$ in the same way (e.g. $x'=x-t$ and $t'=t$). When scaling factors are different, like in (\ref{tr}), then the form of  $I^{(3)}$ in the moving frame differs from the form in the fixed frame, see Appendix~\ref{app1}.}

For those solutions of KdV which preserve their shapes during the motion, that is, for cnoidal solutions and single soliton solutions, integrals of any power of ~$\eta(x,t)$ and any power of arbitrary derivative of the solution with respect to ~$x$ are invariants. That is,
\begin{equation} \label{NZM}
I^{(a,n)} = \int_{-\infty}^{\infty} (\eta_{nx})^a dx =  \mbox{const},
\end{equation}
where ~$n=0,1,2,\ldots$, and ~$a\in \mathbb{R}$ is an arbitrary real number.
Then an arbitrary linear combination of ~$I^{(a,n)}$ is an invariant, as well. 

\subsection{Invariants of the second order equations} \label{niezm2}

Can we construct invariants for KdV type equations of the second order?
Let us try to take $T=\eta$ for  equation  (\ref{etaab}). Then we find that all terms, except ~$\eta_t$, can be written as $X_x$, as
\begin{eqnarray} \label{coneq2}
\hspace{-4ex} & &  \int \left[ 
\eta_x + \alpha\, \frac{3}{2}\eta\eta_x +\beta\,\frac{1}{6} \eta_{3x} 
+ \alpha^2\,\left(-\frac{3}{8}\eta^2\eta_x\right)  \right. \nonumber \\ && \hspace{4ex} \left.
+ \alpha\beta\,\left(\frac{23}{24}\eta_x\eta_{2x}+\frac{5}{12}\eta\eta_{3x} \right)+\beta^2\,\frac{19}{360}\eta_{5x} \right] dx \nonumber\nonumber \\
&=&\hspace{2ex}
\eta +\frac{3}{4}\alpha \eta^2 + \frac{1}{6}\beta \eta_{2x}
 - \frac{1}{8}\alpha^2\eta^3   \\ && \hspace{4ex}
+  \alpha\beta\left(\frac{13}{48}\eta_x^2+ \frac{5}{12}\eta\eta_{2x}\right) +\frac{19}{360}\beta^2\eta_{4x}. \nonumber
\end{eqnarray}
As (\ref{coneq2}) depends on  ~$\eta$~ and space derivatives and also since all those functions vanish when ~$x\to\pm\infty$, the conservation law for mass (volume)
\begin{equation} \label{coneq3}
\int_{-\infty}^{\infty}\eta(x,t)\,dx = \mbox{const.,}
\end{equation}
holds for the second order equation. 

(Conservation law (\ref{coneq3}) holds for the equation with an uneven bottom, as well.)  

Until now we did not find any other invariants for the secod order equations. 
Moreover, we can show that the integral ~$I^{(2)}$ (\ref{I2}) ~{\bf is no longer} an invariant of the second order KdV equation (\ref{etaab}).

Upon multiplication of equation (\ref{etaab}) by  ~$\eta$  one obtains
\begin{eqnarray} \label{e2a}
0 &=& \frac{\partial }{\partial t}\! \left( \frac{1}{2}\eta^2\right)\! + \frac{\partial }{\partial x}\!\left[\frac{1}{2}\eta^2  +\frac{1}{2} \alpha \eta^3 + \frac{1}{6} \beta \!\left(\!-\frac{1}{2} \eta_x^2+\eta\eta_{2x}\!\right) \right. \nonumber \\  &&  \hspace{3ex}
-\frac{3}{32} \alpha^2 \eta^4
 +\frac{19}{360} \beta^2 \!\left(\!\frac{1}{2}\eta_{xx}^2 -\eta_x\eta_{3x}+\eta\eta_{4x}\!\right)  \\  && \left.\hspace{3ex}
+\frac{5}{12}\alpha\beta\, \eta^2\eta_{2x} \right] 
+ \frac{1}{8} \alpha\beta\, \eta\eta_x\eta_{2x} \;. \nonumber
\end{eqnarray}
The last term in (\ref{e2a}) can not be expressed as  ~$\frac{\partial }{\partial x}X(\eta,\eta_x,\ldots )$. Therefore ~$\int^{+\infty}_{-\infty} \eta^2 dx$~ is not a conserved quantity.


 \section{Energy} \label{energy}

The invariant $I^{(3)}$ 
is usually referred to as the energy invariant. 
Is this really the case?

\subsection{Energy in a fixed frame as calculated from the definition}\label{efix}

The hydrodynamic equations for an incompressible, inviscid fluid, in irrotational motion and under gravity in a fixed frame of reference, lead to a KdV equation of the form
\begin{equation} \label{nkdv}
\tilde{\eta}_{\tilde{t}}+\tilde{\eta}_{\tilde{x}} + \alpha\, \frac{3}{2}\tilde{\eta}\tilde{\eta}_{\tilde{x}}  +\beta\,\frac{1}{6} \tilde{\eta}_{3\tilde{x}}=0.
\end{equation}
We will find the function
\begin{equation} \label{w1}
\tilde{f}_{\tilde{x}} = \tilde{\eta} -  \frac{1}{4}\alpha \tilde{\eta}^2 + \frac{1}{3} \beta \tilde{\eta}_{\tilde{x}\tilde{x}},
\end{equation}
 obtained as a byproduct in derivation of KdV,
useful in what follows. For more details see Appendix~\ref{app2} or \cite[Chapter~5]{EIGR}.
Tildas denote scaled dimensionless quantities.

Now construct the total energy of the fluid from the definition.

The total energy is the sum of potential and  kinetic energy. In our two-dimensional system the energy in original (dimensional coordinates) is
\begin{eqnarray}\label{En}
E= T+V &=& \hspace{2ex}\int^{+\infty}_{-\infty}\left(\int^{h+\eta}_{0} \frac{\rho v^2}{2} dy\right) dx \\ &&  +
\int^{+\infty}_{-\infty}\left( \int^{h+\eta}_{0} \rho g y\, dy\right) dx\,. \nonumber
\end{eqnarray}

Equations (\ref{nkdv}) and (\ref{w1}) are obtained after scaling \cite{BS,KRR,KRI}. We now have dimesionless variables, according to 
\begin{equation}\label{Przeskalowanie}
\tilde{\phi}= \frac{h}{l a \sqrt{gh}}  \phi, \quad  \tilde{x}= \frac{x}{l}, \quad \tilde{\eta}=  \frac{\eta}{a}, \quad \tilde{y}= \frac{y}{h}, \quad \tilde{t}= \frac{t}{l/\sqrt{gh}},
\end{equation} 
and
\begin{equation}\label{EnPotNZm}
V=\rho g h^2 l \int^{+\infty}_{-\infty}\int^{1+\alpha\tilde{\eta}}_{0} \rho \, \tilde{y}\, d\tilde{y}\, d\tilde{x}, 
\end{equation}
\begin{equation}\label{EnKinNZm}
T=\frac{1}{2} \rho g h^2 l \int^{+\infty}_{-\infty}\int^{1+\alpha \tilde{\eta}}_{0} \left( \alpha^2 \tilde{\phi}_{\tilde{x}}^2 + \frac{\alpha^2}{\beta} \tilde{\phi}_{\tilde{y}}^2 \right) \, d\tilde{y}\, d\tilde{x}.
\end{equation}
Note, that the factor in front of the integrals has the dimension of energy.

In the following, we omit signs ~$\sim$, having in mind that we are working in dimensionless variables. 
Integration in  (\ref{EnPotNZm}) with respect to ~$y$ yields
\begin{eqnarray}\label{EpRz1}
V &=& \frac{1}{2} g h^2 l \rho  \int_{-\infty }^{\infty } \left(\alpha^2 \eta^2 + 2 \alpha \eta +1 \right) \, dx \\
&=& \frac{1}{2} g h^2 l \rho \left[ \int_{-\infty }^{\infty } \left(\alpha^2 \eta^2 + 2 \alpha \eta  \right) \, dx + \int_{-\infty }^{\infty } dx \right] . \nonumber
\end{eqnarray}
After renormalization (substraction of constant term  $\int_{-\infty }^{\infty } dx $) one obtains
\begin{equation}\label{EpRz1a}
V = \frac{1}{2} g h^2 l \rho  \int_{-\infty }^{\infty } \left(\alpha^2 \eta^2 + 2 \alpha \eta  \right) \, dx  .
\end{equation}

In kinetic energy we use the velocity potential expressed in the lowest (first) order
\begin{equation} 
\phi_x=f_x-\frac{1}{2}\beta y^2 f_{xxx} \quad \mbox{and} \quad 
\phi_y= - \beta y f_{xx},
\end{equation} 
where $f_x$ was defined in (\ref{w1}). 
Now the bracket in the integral (\ref{EnKinNZm}) is
\begin{equation} \label{nawEk}
 \left( \alpha^2 {\phi_x}^2 + \frac{\alpha^2}{\beta} {\phi_y}^2 \right) = \alpha^2 \left(f_x^2 +\beta y^2(-f_xf_{xxx}+f_{xx}^2) \right).
\end{equation}
Inegration with respect to the vertical  corrdinate ~$y$ gives, up to the same order, 
\begin{eqnarray}\label{EkRz1}
T & =& \frac{1}{2} \rho g h^2 l  \!\! \int^{+\infty}_{-\infty} \!\!\!\! \alpha^2
 \left[ f_x^2 (1+\alpha\eta) \right. \nonumber \\ && \left. \hspace{9ex}
 + \beta(-f_xf_{xxx}+f_{xx}^2) \frac{1}{3} (1+\alpha\eta)^3\right]  dx   \\
& =& \frac{1}{2} \rho g h^2 l \!\! \int^{+\infty}_{-\infty} \!\!\!\! \alpha^2
 \left[ f_x^2 + \alpha f_x^2\eta +\frac{1}{3} \beta \left(f_{xx}^2 -f_x f_{xxx}\right)\right]  dx. \nonumber
\end{eqnarray}
In order to express energy through the elevation funcion ~$\eta$ we use  (\ref{w1}).
We then substitute ~$f_x=\eta$ in terms of the third order and ~$f_x^2=\eta^2-\frac{1}{2}\alpha \eta^3 +\frac{2}{3}\beta \eta \eta_{xx}$~ in terms of the second order 

\begin{eqnarray} \label{eketa}
T  &=& \frac{1}{2} \rho g h^2 l \int^{+\infty}_{-\infty} \alpha^2
 \left[ \left(\eta^2-\frac{1}{2}\alpha \eta^3 +\frac{2}{3}\beta \eta \eta_{xx}\right) \right. \nonumber \\ && \hspace{17ex} \left.+ \alpha \eta^3 + \frac{1}{3} \beta\left( \eta_x^2-\eta\eta_{xx}\right)\right]  dx  \nonumber \\
&=& \frac{1}{2} \rho g h^2 l\; \alpha^2 \left[ \int^{+\infty}_{-\infty} 
 \left( \eta^2+\frac{1}{2}\alpha \eta^3  \right)  dx \right. \\ && \hspace{10ex} \left.  + \int^{+\infty}_{-\infty} 
\frac{1}{3}\beta \left(\eta_x^2+\eta\eta_{xx} \right)  dx\right] .\nonumber
\end{eqnarray}
The last term vanishes as 
\begin{equation}\label{PartInt}
\int^{+\infty}_{-\infty}\!\! \!\!\left(\eta_x^2+\eta\eta_{xx}\right) dx =
 \int^{+\infty}_{-\infty}\!\!\!\!\!\!\eta_x^2 dx + \eta\eta_{x}|^{+\infty}_{-\infty}- \int^{+\infty}_{-\infty}\!\!\!\!\!\!\eta_x^2 dx=0.
\end{equation}
Therefore the total energy in the fixed frame is given by
\begin{eqnarray*}\label{EcRz1}
E_\mathrm{tot} \!&\!=\!&\! T+V = \rho g h^2 l    \int_{-\infty }^{\infty } \left( \alpha \eta + (\alpha \eta)^2 + \frac{1}{4}(\alpha \eta)^3  \right) dx  \hspace{5ex} (39)\\  \!&\!=\!&\!
 \rho g h^2 l \!\left(\!\alpha I^{(1)}\!+\!\alpha^2 I^{(2)} \!+\!\frac{1}{4}\alpha^2 I^{(3)} \! +\!\frac{1}{12}\alpha^2\beta  \int_{-\infty }^{\infty } \!\! \eta_x^2\, dx\! \right)  
\end{eqnarray*}
The energy (39) in a fixed frame of reference {\bf has non invariant form}. The last term in (39) results in small deviations from energy conservation only when $\eta_x$ changes in time in soliton's reference frame, what occurs only during soliton collision. This deviations are discussed and illustrated in Section VI~E.

The result  (\ref{EcRz1}) gives the energy in powers of ~$\eta$ only. The same structure of powers in ~$\eta$ was obtained by the authors of \cite{Kalisch}, who work in dimensional KdV equations (\ref{nieruch}) and (\ref{ruch}). On page 122 they present a non-dimensional energy density $E$ in a frame moving with the velocity $U$. Then, if $U=0$ is set, the  energy density in a fixed frame is proportional to                                                                                                                                                                                                                                                                                                                                                                                                                                                                                                                                                                                                                                                                                                                                                                                                                                                                                                                                                                                                                                                                                        $ \alpha\eta +\alpha^2\eta^2$ as the formula is obtained up to second order in $\alpha$. However, the third order term is $\frac{1}{4}\alpha^3\eta^3$,  so the formula up to the third order  in  $\alpha$ becomes
\begin{equation} \label{KalEdn} 
E\sim \alpha\eta +\alpha^2\eta^2 +\frac{1}{4}\alpha^3\eta^3.
\end{equation}
This energy density contains the same  terms like  (\ref{EcRz1}) and does not contain the term $\eta_x^2$, as well.

Energy calculated from the definition does not contain a proper invariant of motion. 

\subsection{Energy in a moving frame }

Now consider the total energy according to (28) calculated in a frame moving with the velocity of sound $c=\sqrt{gh}$. Using the same scaling (29) to dimensionless variables we note that in these variables $c=1$.
As pointed by Ali and Kalisch [8,Sect.~3] 
 working in such system one has to replace
$\phi_x$ by the horizontal velocity in a moving frame, that is by $\tilde{\phi}_{\tilde{x}}
-\frac{1}{\alpha}=\alpha\tilde{\eta} -  \frac{1}{4}\alpha \tilde{\eta}^2 + \beta\left(\frac{1}{3} -\frac{y^2}{2}\right) \tilde{\eta}_{\tilde{x}\tilde{x}}-\frac{1}{\alpha}$.
Then repeating the same steps as in the previous subsection yields the energy expressed by invariants 
\begin{eqnarray}\label{ETmov}
E_\mathrm{tot}\! &\!=\!&\!\rho g h^2 l \! \!  \int_{-\infty }^{\infty }\! \left[-\frac{1}{2}\alpha 	\tilde{\eta}\! +\!\frac{1}{4} (\alpha \tilde{\eta})^2 \!+\! \frac{1}{2}\alpha^3\!\left(\!\tilde{\eta}^3\!-\! \frac{1}{3} \frac{\beta}{\alpha}\tilde{\eta}_{\tilde{x}}^2\!\right)\! \right]\! d\tilde{x}  \nonumber \\
 &=& \rho g h^2 l \left(-\frac{1}{2} \alpha I^{(1)} + \frac{1}{4} \alpha^2 I^{(2)} + \frac{1}{2} \alpha^3 I^{(3)} \right).
\end{eqnarray}
The crucial term ~$-\frac{1}{6}\alpha^2\beta\,\tilde{\eta}_{\tilde{x}}^2$ in (\ref{ETmov}) appears due to integration over vertical variable of the term ~$\frac{\beta}{\alpha}\tilde{\eta}_{\tilde{x}\tilde{x}}$ supplied by $(\tilde{\phi}_{\tilde{x}}-\frac{1}{\alpha})^2$.

\section{Variational approach} \label{varia}

\subsection{Lagrangian approach, potential formulation }

Some attempts at the variational approach to shallow water problems are summarized in 
 G.B.~Whitham's book \cite[Sect~16.14]{Whit}.

For KdV as it stands, 
we can not write a variational principle directly. It is necessary to introduce a velocity potential. The simplest choice is to take ~$\eta=\varphi_x$.
Then  equation (\ref{kdv1}) in the fixed frame takes the form
\begin{equation} \label{e1p}
\varphi_{xt} + \varphi_{xx}+ \frac{3}{2}\alpha \varphi_x \varphi_{xx}+ \frac{1}{6}\beta\varphi_{xxxx}=0\,.
\end{equation}
The appropriate Lagrangian density is
\begin{equation} \label{L1Wn} 
\mathcal{L}_\mathrm{fixed~frame}  :=-\frac{1}{2} \varphi_t\varphi_x -\frac{1}{2}\varphi_{x}^2 -\frac{\alpha}{4}\varphi_x^3+\frac{\beta}{12} \varphi_{xx}^2 \,  .
\end{equation}
 Indeed, the Euler--Lagrange equation obtained from Lagrangian (\ref{L1Wn}) is just (\ref{e1p}).

For our moving reference frame the substitution  ~$\eta=\varphi_x$ into (\ref{kdv1m}) gives
\begin{equation} \label{e1pr}
\varphi_{xt} + \frac{3}{2}\alpha \varphi_x \varphi_{xx}+ \frac{1}{6}\beta\varphi_{xxxx}=0\,.
\end{equation}
So, the appropriate Lagrangian density is
\begin{equation} \label{L1Wr} 
\mathcal{L}_\mathrm{moving~frame}  :=-\frac{1}{2} \varphi_t\varphi_x  -\frac{\alpha}{4}\varphi_x^3+\frac{\beta}{12} \varphi_{xx}^2 \,  .
\end{equation}
Again, the Euler--Lagrange equation obtained from Lagrangian (\ref{L1Wr}) is just (\ref{e1pr}).


\subsection{Hamiltonians for KdV equations in the potential formulation }

The Hamiltonian for the KdV equation in a fixed frame  (\ref{kdv1}) can be obtained in the following way. Defining generalized momentum ~$\displaystyle\pi=\frac{\partial  \mathcal{L}}{\partial \varphi_{t}}$, where ~$\mathcal{L}$ is given by (\ref{L1Wn}),
one obtains
\begin{eqnarray} \label{Hh1}
H &=& \int_{-\infty}^{\infty} \left[\pi \dot{\varphi}-\mathcal{L}\right] dx=  \int_{-\infty}^{\infty} \left[\frac{1}{2} \varphi_{x}^2 +\frac{\alpha}{4}\varphi_{x}^3-\frac{\beta}{12} \varphi_{xx}^2\right] dx \nonumber \\
&=& \int_{-\infty}^{\infty} \left[\frac{1}{2}\eta^2+\frac{1}{4}\alpha\left( \eta^3-\frac{\beta}{3\alpha} \eta_{x}^2\right)\right] dx \,.
\end{eqnarray}
The energy is expressed by invariants ~$I^{(2)}, I^{(3)}$ so it is a constant of motion.

The same procedure for  KdV  in a moving frame  (\ref{kdv1m}) 
gives
\begin{eqnarray} \label{Hh2}
H &=&\int_{-\infty}^{\infty} \left[ \pi \dot{\varphi}-\mathcal{L}\right] dx=\int_{-\infty}^{\infty} \left[ \frac{\alpha}{4} \varphi_{x}^3-\frac{\beta}{12} \varphi_{xx}^2\right] dx \nonumber \\
&=& \frac{1}{4}\alpha \int_{-\infty}^{\infty}\left(\eta^3-\frac{\beta}{3\alpha} \eta_{x}^2\right) dx \,.
\end{eqnarray}
The Hamiltonian (\ref{Hh2})  only consists ~$I^{(3)}$.

 The constant of motion in a moving frame is
\begin{equation} \label{H2C}
E = \frac{1}{4}I^{(3)} = \mbox{const}.
\end{equation}

The potential formulation of the Lagrangian, described above, is succesful for deriving  KdV equations both for fixed and moving reference frames. It fails, however, for the second order KdV equation~(\ref{etaab}). We proved that there exists a nonlinear expression of $\mathcal{L}(\varphi_t,\varphi_x,\varphi_{xx},\ldots)$, such that the resulting Euler--Lagrange equation differs very little from equation (\ref{etaab}). The difference lies only in the value of one of the coefficients in the second order term
$~\alpha\beta\,\left(\frac{23}{24}\eta_x\eta_{2x}+\frac{5}{12}\eta\eta_{3x} \right)$. 
Particular values of coefficients in this term also cause the lack of the ~$I^{(2)}$ invariant for second order KdV equation,  (see (\ref{e2a})).

\section{Luke's Lagrangian and KdV energy}\label{LukL}

The full set of Euler equations for the shallow water problem, as well as KdV equations (\ref{kdv1}), (\ref{kdvm}), and second order KdV equation (\ref{etaab}) can be derived from Luke’s Lagrangian \cite{Luke}, see, e.g. \cite{MS}. Luke points out,
however, that his Lagrangian is not equal to the difference of kinetic and potential energy. Euler–Lagrange equations obtained from $L=T-V$ do not have the proper form at the boundary. Instead, Luke’s Lagrangian  is the sum of kinetic and potential energy suplemented by the  $\phi_t$ term which is necessary in dynamical
boundary condition.

\subsection{Derivation of KdV energy from the original Euler equations according
to \cite{EIGR}}\label{IRbook}

In Chapter 5.2 of the Infeld and Rowlands book the authors present a derivation of the KdV equation from the Euler (hydrodynamic) equations using a single small parameter ~$\varepsilon$. Moreover, they show that the same method allows us to derive the Kadomtsev--Petviashvili (KP) equation \cite{KP}  for water waves \cite{IRH,KP_woda,L-H&F,Benj} and also nonlinear equations for ion acoustic waves in a plasma \cite{KP_plazma,InRo}. The authors first derive equations of motion, then construct a Lagrangian and look for constants of motion.
For the purpose of this paper and for comparison to results obtained  in the next subsections it is convenient
 to present their results starting from Luke's Lagrangian density. 
That density  can be written as (here $g=1$)\begin{equation}\label{LLuke}L=\int_0^{1+\eta} \left[\phi_t +\frac{1}{2}(\phi_x^2+\phi_z^2) +z \right] dz  \,.\end{equation}

In Chapter 5.2.1 of \cite{EIGR} the authors introduce scaled variables in a movimg frame ($\varepsilon$ plays a role of small parameter and if ~$\varepsilon=\alpha=\beta$, then KdV equation is obtained).
Then (for details, see Appendix~\ref{app2} or \cite[Chapter~5.2]{EIGR})
\begin{eqnarray}\label{5.2.20}
\phi_z &=& -\varepsilon^{\frac{3}{2}} z \Green{f_{\xi\xi}}, \qquad \phi_x=  \varepsilon f_\xi-\varepsilon^2\frac{z^2}{2}\Green{f_{\xi\xi\xi}}, \nonumber\\
 \phi_t &=& - \varepsilon f_\xi+\varepsilon^2\left(f_\tau+\frac{z^2}{2}\Green{f_{\xi\xi\xi}} \right) -\varepsilon^3\frac{z^2}{2}f_{\xi\xi\tau}.
\end{eqnarray}
Substitution of the above formulas into the expression [~]  under the integral in (\ref{LLuke}) gives
\begin{eqnarray}\label{[(1)]}
[~] &=& z - \varepsilon f_\xi + \varepsilon^2\left(f_\tau+\frac{1}{2}f_\xi^2+ \frac{z^2}{2}f_{\xi\xi\xi} \right)  \\
&& \hspace{2ex} + \varepsilon^3\frac{z^2}{2}\left[-f_{\xi\xi\tau} + (f_{\xi\xi}^2-f_\xi f_{\xi\xi\xi} )\right] +O(\varepsilon^4).\nonumber
\end{eqnarray}
{\bf Remark}~ 
The full Lagrangian is obtained by integration of the Lagrangian density  (\ref{LLuke}) with respect to ~$x$. Integration limits are ~$(-\infty,\infty)$ for a soliton solutions, or ~$[a,b]$, where ~$b-a=X$--wave length (space period) for cnoidal solutions. Integration by parts and properties of the solutions at the  limits, see (\ref{limits}), allow us to use the equivalence   $\int_{-\infty}^{\infty}(f_{\xi\xi}^2-f_\xi f_{\xi\xi\xi})d\xi = \int_{-\infty}^{\infty} 2f_{\xi\xi}^2 d\xi$.

Therefore 
\begin{eqnarray}\label{[(1)a]}
[~] &=& z - \varepsilon f_\xi + \varepsilon^2\left(f_\tau+\frac{1}{2}f_\xi^2+ \frac{z^2}{2}f_{\xi\xi\xi} \right) \\ && \hspace{2ex} 
+ \varepsilon^3\frac{z^2}{2}\left[-f_{\xi\xi\tau} + 2f_{\xi\xi}^2\right] +O(\varepsilon^4).\nonumber
\end{eqnarray}
Integration over $y$ gives (note that $1+\eta\Longrightarrow 1+\varepsilon\eta$)
\begin{eqnarray}\label{Luke} 
L &=& \frac{1}{2}  (1+\varepsilon\eta)^2 +  (1+\varepsilon\eta)\left[- \varepsilon f_\xi + \varepsilon^2\left(f_\tau +\frac{1}{2} f_\xi^2 \right)\right] \nonumber \\ &&  
+\frac{1}{3}(1+\varepsilon\eta)^3 \left[\frac{1}{2}\varepsilon^2 f_{\xi\xi\xi} -\frac{1}{2}\varepsilon^3  f_{\xi\xi\tau} +\varepsilon^3 f_{\xi\xi}^2  \right].
\end{eqnarray}
Write (\ref{Luke}) up to third order in ~$\varepsilon$
$$L= L^{(0)}+\varepsilon  L^{(1)} + \varepsilon^2  L^{(2)} +\varepsilon^3  L^{(3)} +O (\varepsilon^4) \,.$$
It is easy to show, that
\begin{eqnarray}\label{L03} 
L^{(0)} & = & \frac{1}{2}, \qquad
L^{(1)}  =  \eta -f_{\xi}, \nonumber\\
L^{(2)} & = & f_\tau+\frac{1}{2} \eta^2 -\eta f_{\xi}+ \frac{1}{2}f_{\xi}^2 +  \frac{1}{6}f_{\xi\xi\xi}, \\
L^{(3)} & = & \eta f_{\tau} +  \frac{1}{2}\eta f_{\xi}^2+\frac{1}{2}\eta f_{\xi\xi\xi}- \frac{1}{6}f_{\xi\xi\tau} + \frac{1}{3}f_{\xi\xi}^2.  \nonumber 
\end{eqnarray}
The Hamiltonian density reads as 
\begin{eqnarray} \label{Hden} 
H &=& f_\tau \frac{\partial L}{\partial f_\tau } + f_{\xi\xi\tau}  \frac{\partial L}{\partial f_{\xi\xi\tau} } -L \\
& =& -\left[\frac{1}{2} +\varepsilon \left(\eta -f_{\xi}
\right) +\varepsilon^2 \left( \frac{1}{2} \eta^2 -\eta f_{\xi}+ \frac{1}{2}f_{\xi}^2 +  \frac{1}{6}f_{\xi\xi\xi}\right)  \right. \nonumber\\
& & \left. \hspace{3ex }+ \varepsilon^3 \left(  \frac{1}{2}\eta f_{\xi}^2+\frac{1}{2}\eta f_{\xi\xi\xi} + \frac{1}{3}f_{\xi\xi}^2\right)\right]. \nonumber
\end{eqnarray} 
Dropping the constant term one obtains the total energy as
\begin{eqnarray} \label{H03} 
\mathcal{E} &=& \int_{-\infty}^{\infty} \left[\varepsilon \left(\eta -f_{\xi}
\right) +\varepsilon^2 \left( \frac{1}{2} \eta^2 -\eta f_{\xi}+ \frac{1}{2}f_{\xi}^2 +  \frac{1}{6}f_{\xi\xi\xi}\right) \right. \nonumber \\ && \left. \hspace{5ex }
 + \varepsilon^3 \left(  \frac{1}{2}\eta f_{\xi}^2+\frac{1}{2}\eta f_{\xi\xi\xi} + \frac{1}{3}f_{\xi\xi}^2\right)\right]d\xi.
\end{eqnarray}

Now, we need to express $f_\xi$ and its derivatives by $\eta$ and its derivatives. We use (\ref{w1}) replacing $\alpha$ and $\beta$ by $\varepsilon$, that is,
\begin{equation} \label{w2}
f_{\xi}=\eta - \frac{1}{4} \varepsilon \eta^2 + \frac{1}{3}\varepsilon \eta_{\xi\xi}.
\end{equation}

Then the total energy in a moving frame is expressed in terms of the second and the third  invariants
\begin{equation} \label{H03b} 
\mathcal{E} =-\left[\varepsilon^2 \frac{1}{4}\int_{-\infty}^{\infty} \!\!\!
  \eta^2 \, dx +\varepsilon^3 \frac{1}{2}\int_{-\infty}^{\infty}  \!\!\!
\left(  \eta^3 -\frac{1}{3}\eta_\xi^2 \right) dx \right].
\end{equation}

 Note that the term ~$\frac{1}{3}\eta_\xi^2$~ occuring in the third order invariant originates from three terms appearing in ~$\phi_z^2$,  ~$\phi_x^2$~ and $\phi_t$ (see terms ~$f_{\xi\xi}$~ and ~$f_{\xi\xi\xi}$~ in (\ref{5.2.20})).

\subsection{Luke's Lagrangian} 

The original Lagrangian density in  Luke's paper \cite{Luke} is
\begin{equation}\label{LLu1}
L=\int_0^{h(x)} \rho \left[\phi_t +\frac{1}{2}(\phi_x^2+\phi_y^2) +gy \right] dy  \,.
\end{equation}
After scaling as in \cite{BS,KRR,KRI}
\begin{equation}\label{scal}
 \tilde{\phi}= \frac{h}{la\sqrt{gh}} \phi, \quad \tilde{x} =\frac{x}{l},  \quad 
\tilde{\eta} = \frac{\eta}{a}, \quad \tilde{y} =\frac{y}{h}, \quad \tilde{t} = \frac{t}{l/\sqrt{gh}},
\end{equation}
we obtain
\begin{equation}\label{sc1}
\phi_t =gh\alpha \,\tilde{\phi}_{\tilde{t}},\qquad \phi_x^2 = gh\alpha^2 \,\tilde{\phi}_{\tilde{x}}^2, \qquad \phi_y^2 = gh \frac{\alpha^2}{\beta} \,\tilde{\phi}_{\tilde{y}}^2.
\end{equation}
The  Lagrangian density in scaled variables becomes
($dy = h d \tilde{y}$)
\begin{eqnarray} \label{LL2}
L &=& \rho g h a \int_0^{1+\alpha\eta} \left[ \tilde{\phi}_{\tilde{t}} +\frac{1}{2} \left( \tilde{\phi}_{\tilde{x}}^2+\frac{\alpha^2}{\beta} \,\tilde{\phi}_{\tilde{y}}^2 \right)\right]d\tilde{y}\nonumber \\ && 
+ \frac{1}{2}  \rho g h^2 (1+\alpha\eta)^2 . 
\end{eqnarray}
So, in dimensionless quatities 
\begin{equation} \label{LL}
\frac{L}{\rho g h a}  =  \int_0^{1+\alpha\eta} \left[ \tilde{\phi}_{\tilde{t}} +\frac{1}{2} \left(\alpha \tilde{\phi}_{\tilde{x}}^2+\frac{\alpha}{\beta} \,\tilde{\phi}_{\tilde{y}}^2 \right)\right]d\tilde{y} + \frac{1}{2} \alpha\eta^2,
\end{equation}
where the constant term and the term proportional to ~$\eta$ in the expansion of ~$(1+\alpha\eta)^2$ are omitted. The form (\ref{LL}) is identical with  Eq.~(2.9) in Marchant \& Smyth  \cite{MS}.

The full Lagrangian is obtained by integration over ~$x$. In dimensionless variables ($dx= l\,d\tilde{x}$) it gives 
\begin{equation} \label{LLen}
\mathcal{L}\! =\! E_0 \!\int_{-\infty}^{\infty}\! \left[\! \int_0^{1+\alpha\eta}\! \left[ \tilde{\phi}_{\tilde{t}} +\frac{1}{2} \left(\alpha \tilde{\phi}_{\tilde{x}}^2+\frac{\alpha}{\beta} \,\tilde{\phi}_{\tilde{y}}^2 \right)\!\right]d\tilde{y}\! +\! \frac{1}{2} \alpha\eta^2 \right]\! d\tilde{x}.
\end{equation}
The factor in front of the integral, ~$E_0=\rho g h a l = \rho g h^2 l\,\alpha$, has the dimension of energy.

Next, the signs ( $\sim$ ) will be omitted, but we have to remember that we are working in scaled dimensionless variables in a fixed reference frame.

\subsection{Energy in the fixed reference frame} 

Express the Lagrangian density by ~$\eta$~ and ~$f=\phi^{(0)}$.
Now, up to  first order in small parameters
\begin{eqnarray} \label{pot}
\phi &=& f -\frac{1}{2} \beta y^2 f_{xx},\qquad \phi_t=f_t - \frac{1}{2} \beta y^2 f_{xxt}, \nonumber \\
\phi_x &=& f_x - \frac{1}{2} \beta y^2 f_{xxx},  \qquad 
\phi_y = -  \beta y f_{xx}.
\end{eqnarray}
Then the expression under the integral in (\ref{LL}) becomes
\begin{equation} \label{pot2}
[~] = f_t - \frac{1}{2} \beta y^2 f_{xxt} + \frac{1}{2} \alpha f_x^2 + \frac{1}{2} \alpha\beta y^2 \left(- f_x f_{xxx} + f_{xx}^2\right).
\end{equation}
From properties of solutions at the limits ~$\left(- f_x f_{xxx} + f_{xx}^2\right) \Rightarrow 2 f_{xx}^2$. 
Integration of (\ref{pot2}) over~$y$ yields
\begin{eqnarray} \label{LLn}
\frac{L}{\rho g h a}  &=& \left( f_t + \frac{1}{2} \alpha f_x^2 \right)(1+\alpha\eta) - \frac{1}{2} \beta f_{xxt}\,\frac{1}{3} (1+\alpha\eta)^3  \nonumber \\ &&
 + \alpha\beta f_{xx}^2\, \frac{1}{3} (1+\alpha\eta)^3 +\frac{1}{2}\alpha\eta^2.
\end{eqnarray}

The dimensionless Hamiltonian  density is\\ ($f_t \frac{\partial L}{\partial f_t}+f_{xxt}  \frac{\partial L}{\partial f_{xxt}}-L$)
\begin{equation} \label{H02}
\frac{H}{\rho g h^2 l} \!= \!-\alpha\! \left[ \frac{1}{2} \alpha f_x^2(1+\alpha\eta)+ \alpha\beta f_{xx}^2\, \frac{1}{3} (1+\alpha\eta)^3 +\frac{1}{2}\alpha\eta^2 \right]\!.
\end{equation}
Again, we need to express the Hamiltonian by ~$\eta$ and its derivatives, only.
Inserting
\begin{equation} \label{fx2}
f_x=\eta- \frac{1}{4}\alpha\eta^2+\frac{1}{3}\beta\eta_{xx}
\end{equation}
 into (\ref{H02}) and leaving  terms up to third order one obtains 
\begin{equation} \label{H0}
\frac{H}{\rho g h^2 l} =-\alpha \left[ \alpha \eta^2+\frac{1}{4} \alpha^2 \eta^3 +\frac{1}{3}\alpha\beta(\eta_x^2+\eta\eta_{xx})  \right].
\end{equation}
The energy is
\begin{eqnarray} \label{H0ee}
\frac{E}{\rho g h^2 l} &=& -\alpha\! \int_{-\infty}^{\infty}\left[ \alpha \eta^2+\frac{1}{4} \alpha^2 \eta^3 +\frac{1}{3}\alpha\beta(\eta_x^2+\eta\eta_{xx})  \right] dx \nonumber \\ &=&
 -\left[ \alpha^2  \! \int_{-\infty}^{\infty} \!\!\eta^2 dx +\frac{1}{4} \alpha^3 \!\int_{-\infty}^{\infty} \!\!\eta^3 dx \right] 
\end{eqnarray}
since the integral of the $\alpha\beta$ term vanishes. Here, in the same way as in calculations of energy directly from the definition (\ref{EcRz1}), the energy is expressed by integrals of $\eta^2$ and $\eta^3$. The term proportional to $\alpha\eta$ is not present in (\ref{H0ee}), because it was dropped  earlier \cite{MS}.

\subsection{Energy in a moving frame}
Transforming into the moving frame
\begin{equation} \label{ur}
\bar{x}=x-t, \quad \bar{t}=\alpha t, \quad \partial_x = \partial_{\bar{x}},
\quad \partial_t = - \partial_{\bar{x}}+ \alpha \partial_{\bar{t}}.
\end{equation} 
\begin{equation} \label{ur1}
\phi = f -\frac{1}{2} \beta y^2 f_{\bar{x}\bar{x}}, \quad \phi_x=f_{\bar{x}} -\frac{1}{2} \beta y^2 f_{\bar{x}\bar{x}\bar{x}}, \quad \phi_y = -\beta y f_{\bar{x}\bar{x}},
\end{equation}
\begin{equation} \label{ur2}
\phi_t= -f_{\bar{x}} + \frac{1}{2} \beta y^2 f_{\bar{x}\bar{x}\bar{x}} +\alpha (f_{\bar{t}}- \frac{1}{2} \beta y^2 f_{\bar{x}\bar{x}\bar{t}}).
\end{equation}
Up to second order 
\begin{eqnarray} \label{ur3}
\frac{1}{2} \left(\alpha \phi_x^2+\frac{\alpha}{\beta} \phi_y^2\right)&=&\frac{1}{2} \left[\alpha f_{\bar{x}}^2 + \alpha\beta y^2 (-f_{\bar{x}}f_{\bar{x}\bar{x}\bar{x}}+ f_{\bar{x}\bar{x}}^2) \right] \nonumber \\
 &=& \frac{1}{2} \alpha f_{\bar{x}}^2 + \alpha\beta y^2 f_{\bar{x}\bar{x}}^2.
\end{eqnarray}
Therefore the expression under the integral in  (\ref{LL}) is
\begin{equation} \label{ur4}
[~~] =  -f_{\bar{x}} + \frac{1}{2} \beta y^2 f_{\bar{x}\bar{x}\bar{x}} +\alpha (f_{\bar{t}}- \frac{1}{2} \beta y^2 f_{\bar{x}\bar{x}\bar{t}}) +\frac{1}{2} \alpha f_{\bar{x}}^2 + \alpha\beta y^2 f_{\bar{x}\bar{x}}^2.
\end{equation}

Integration yields
\begin{eqnarray} \label{LLu}
\frac{L}{\rho g h a} \! & = &\!
\left( -f_{\bar{x}} + \alpha f_{\bar{t}} +\frac{1}{2} \alpha f_{\bar{x}}^2 \right) (1+\alpha\eta)   \\ \!& + &\!
\frac{1}{3} (1+\alpha\eta)^3 \left(\!\frac{1}{2} \beta (f_{\bar{x}\bar{x}\bar{x}} -f_{\bar{x}\bar{x}\bar{t}}) +\alpha\beta f_{\bar{x}\bar{x}}^2\! \right)\! + \!\frac{1}{2}\alpha \eta^2.\nonumber
\end{eqnarray}

Like in (\ref{H02}) above, the Hamiltonian density is 
\begin{eqnarray} \label{hhuu}
\frac{H}{\rho g h^2 l} & = & -\alpha \left[
\left( -f_{\bar{x}}  +\frac{1}{2} \alpha f_{\bar{x}}^2 \right) (1+\alpha\eta)\right. \\
&& + \left.\frac{1}{3} (1+\alpha\eta)^3 \left(\frac{1}{2} \beta f_{\bar{x}\bar{x}\bar{x}}  +\alpha\beta f_{\bar{x}\bar{x}}^2 \right) + \frac{1}{2}\alpha \eta^2 \right]. \nonumber
\end{eqnarray}
Expressing $f_{\bar{x}}$ by (\ref{fx2}) one obtains
\begin{eqnarray} \label{hhuu1}
\frac{H}{\rho g h^2 l} & = & -\alpha \left[ -\frac{1}{4}\alpha \eta^2+\frac{1}{3}\beta \eta_{xx} -\frac{1}{2}\alpha^2 \eta^3\right. \\ &&
+ \left. \alpha\beta\left( -\frac{1}{4} \eta_x^2-\frac{5}{12} \eta\eta_{xx} \right)-\frac{1}{18} \beta^2 \eta_{xxxx}\right]. \nonumber
\end{eqnarray}
Finally the energy is given by 
\begin{equation} \label{hhuu2}
\frac{E}{\rho g h^2 l} = \alpha^2\frac{1}{4} \int_{-\infty}^{\infty} \eta^2 dx +\alpha^3 \frac{1}{2}  \int_{-\infty}^{\infty} \left( \eta^3 -\frac{1}{3} \frac{\beta}{\alpha} \eta_x^2\right) dx
\end{equation}
since integrals from terms with ~$\beta,\beta^2$~ vanish at integration limits, and ~$-\frac{5}{12} \eta\eta_{xx}\Rightarrow \frac{5}{12}\eta_x^2$.
 The invariant term proportional to $\alpha\eta$ is not present in (\ref{hhuu2}), because it was dropped in (\ref{LL}). If we include that term, the total energy is a linear combination of all three lowest invariants, $I^{(1)}, I^{(3)},I^{(3)}$.\\

\noindent{\bf Comment~~} {\sl An almost identical formula for the energy in a moving frame, for KdV  expressed in dimensional variables (\ref{ruch}), was obtained in \cite{Kalisch}. 
That energy is expressed by all three lowest order invariants
\begin{eqnarray} \label{kal3}  \mathcal{E}&=&-\frac{1}{2} c^2\int_{-\infty}^{\infty} \eta\,dx + \frac{1}{4} \frac{c^2}{h}\int_{-\infty}^{\infty} \eta^2\,dx \\
&&+ \frac{1}{2} \frac{c^2}{h^2}\int_{-\infty}^{\infty} \left( \eta^3-\frac{h^3}{3}\eta_x^2 \right)dx, \nonumber  \end{eqnarray} as well.
Translation of (\ref{kal3}) to nondimensional variables yields
$$ \mathcal{E}\varrho = \varrho g h^2 l \left(-\frac{1}{2} \alpha I^{(1)} + \frac{1}{4} \alpha^2 I^{(2)} + \frac{1}{2} \alpha^3 I^{(3)} \right). $$
}

\subsection{How strongly is energy conservation violated?}

\begin{figure}[tbh]
\begin{center}
\resizebox{1.0\columnwidth}{!}{\includegraphics{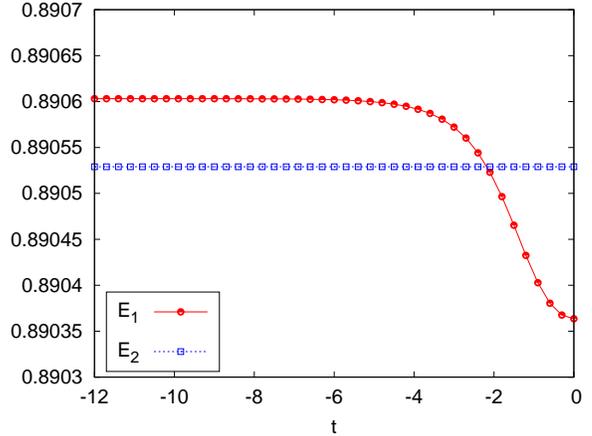}}
\end{center}
\vspace{-5mm}
\caption{Precision of energy conservation for 3-soliton solution.  Energies are plotted as open circles ($E_1$) and open squares ($E_2$) for 40 time instants.} 
 \label{s3inv}
\end{figure}

The total energy in the fixed frame is given by equation~(\ref{EcRz1}). Taking into account its non-dimensional part we may write
\begin{eqnarray} \label{Enoncons}
E_1(t) &=& \frac{T+V}{\varrho g h^2 l }=\int_{-\infty}^{\infty}\left[\alpha\eta+(\alpha\eta)^2 + \frac{1}{4}(\alpha\eta)^3 \right] dx \nonumber \\
&=& \alpha I^{(1)} +\alpha^2 I^{(2)}+\frac{1}{4} \int_{-\infty}^{\infty}(\alpha\eta)^3 dx
\end{eqnarray}
In order to see how much the changes of $E_1$ violate energy conservation we will compare it to the same formula but expressed by invariants
\begin{equation} \label{Econs}
E_2(t) =  \alpha I^{(1)} +\alpha^2  I^{(2)}+\frac{1}{4}\alpha^3 I^{(3)}.
\end{equation}

The time dependence of $E_1$ and $E_2$ is presented in Fig.~\ref{s3inv} for a 3-soliton solution of KdV  (\ref{kdv1}).  Presented is time evolution in the interval $t\in[-12,0]$. The shape of the 3-soliton solution is presented only for three times $t=-12,-6,0$ in order to show shapes changing during the collision. 

\begin{figure}[tbh]
\begin{center}
\resizebox{1.0\columnwidth}{!}{\includegraphics{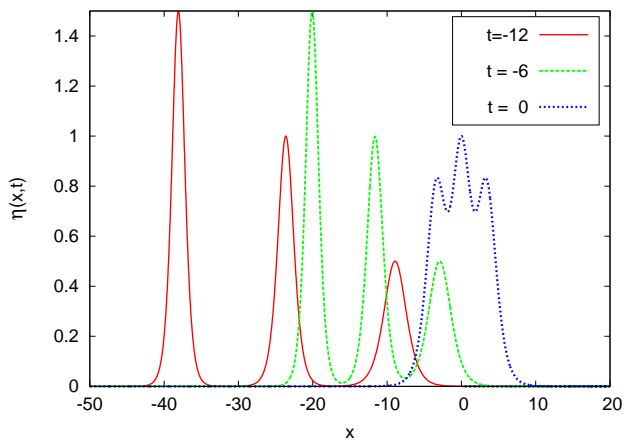}}
\end{center}
\vspace{-5mm}
\caption{Shape evolutiom of 3-soliton solution during collision.} 
 \label{s3ev}
\end{figure}

For presentation the example of a 3-soliton solution with amplitudes equal 1,5, 1 and 0.5 was chosen. In Fig.~\ref{s3ev} the positions of solutions at given times were artificially shifted to set them closer to each other. The plots in Figs.~\ref{s3inv} and \ref{s3ev} for $t>0$ are symmetric to those which are shown in the figures.

For this example the relative discrepancy of the enregy $E_1$ from the constant value, is very small 
\begin{equation} \label{Enonc}
\delta E = \frac{E_1(t=-12) -E_1(t=0)}{E_1(t=-12)} \approx 0.000258.
\end{equation}
However, the $E_2$ energy is conserved with numerical precision of thirteen decimal digits in this example. 
In a similar example with a 2-soliton solution (with apmlitudes 1 and 0.5) the relative error (\ref{Enonc}) was even smaller, with the value $\delta E\approx 0.00014$. This suggests that the degree of nonconservation of energy increases with $n$, where $n$ is the number of solitons in the solution. 

\begin{table*}[tb]
\caption{\label{tab}Comparison of different energy formulas. Here ~$\displaystyle \eta^{(3)}= \int_{-\infty}^{\infty} \eta^3 \,dx $. 
$~^\dagger$ Formulas in this column are written in $\displaystyle \frac{E}{\varrho g h^2 l}$.} 
\begin{ruledtabular}
\begin{tabular}{|l|c|c|c|c|c|}
& Euler & Luke's & Integrals & Potential & KdV \\ 
& equations & Lagrangian & $T+V$ & Lagrangian & dimensional $^\dagger$\\
 & & & & & \\
 \hline
 & &  & & & \\
Fixed frame & $\!\alpha I^{(1)}\!+\!\alpha^2 I^{(2)}\!+\!\frac{1}{4}\alpha^3 \eta^{(3)}\!$ & $\!\alpha^2 I^{(2)}\!+\!\frac{1}{4} \alpha^3 \eta^{(3)}\!$ & $\!\alpha I^{(1)}\!+\!\alpha^2 I^{(2)}\!+\!\frac{1}{4}\alpha^3  \eta^{(3)}\!$ & $\!\frac{1}{2}I^{(2)}\!+\!\frac{1}{4}\alpha I^{(3)}\!$ & $\!\alpha I^{(1)} \!+\!\alpha^2 I^{(2)}\!+\!\frac{1}{4}\alpha^3 \eta^{(3)}\!$ \\ 
 & (\ref{KalEdn}) &  (\ref{H0ee}) &(\ref{EcRz1}) & (\ref{Hh1}) &  (\ref{KalEdn})\\
\hline
 & & & & & \\
Moving frame & $\!\frac{1}{4}\alpha^2 I^{(2)}\!+\!\frac{1}{2}\alpha^3 I^{(3)}\!$ & $\!\frac{1}{4}\alpha^2 I^{(2)}\!+\!\frac{1}{2}\alpha^3I^{(3)}\!$ &  
$\!-\frac{1}{2}\alpha I^{(1)}\!+\!\frac{1}{4}\alpha^2 I^{(2)}\!+\!\frac{1}{2}\alpha^3 I^{(3)}\!$ 
& $\!\frac{1}{4}\alpha I^{(3)}\!$ & $\!-\frac{1}{2}\alpha I^{(1)}\!+\! \frac{1}{4} \alpha I^{(2)}\!+\!\frac{1}{2}\alpha^2 I^{(3)}\!$ \\
 & (\ref{H03b}) &  (\ref{hhuu2}) & (\ref{ETmov})  &(\ref{Hh2}) & (\ref{kal3})\\
\end{tabular}
\end{ruledtabular}
\end{table*}

\section{Conclusions for KdV equation}\label{concl}

The main conclusions can be formulated as follows
\begin{itemize}
 \item The invariants of KdV in  fixed and  moving frames have the same form. (Of course when we have the same scaling factor for $x$~ and $t$ in the transformation between frames).
 \item We confirmed some known facts. Firstly, that
the usual form of the energy ~$H=T+V$~ is not always expressed by invariants only. The reason lies in the fact, as pointed out by Luke in \cite{Luke}, that the Euler--Lagrange equations obtained from the Lagrangian ~$L=T-V$~ do not supply the right boundary conditions.
Secondly, the 
variational approach based on Luke's Lagrangian density provides the right Euler equations at the boundary and allows for a derivation higher order KdV equations.

\item In the frame moving  with the velocity of sound all energy components are expressed by invariants. Energy is conserved. 

\item Numerical calculations confirm that invariants ~$I^{(1)},I^{(2)},I^{(3)}$~ in the forms (\ref{I1}), (\ref{I2}), (\ref{i3E}), (\ref{i3Em}) are exact constants of motion for two- and three-soliton solutions, both for fixed and moving coordinate systems. In all performed tests the invariants were exact up to fourteen digits in double precision calculations.

\item For the extended KdV equation (\ref{etaab}) we have only found one invariant of motion $I^{(1)}$  (\ref{coneq3}).

\item The total energy in the fixed coordinate system as calculated in (\ref{EcRz1}) is not exactly conserved but only altered during collisions, even then by minute quantities (an order of magnitude smaller than expected). Details in figure caption of figure \ref{s3inv}.

\end{itemize}

A summary of these conclusions can be found in Table~\ref{tab}.


\section{Extended KdV equation} \label{extended}

In this section we calculate energy formula coresponding to a wave motion governed by second order equations in scaled variables, that is the equation (\ref{etaab}) for the fixed coordinate system and the correponding equation  for a moving coordinate system. As previosly we compare energies calculated from the definition with those Luke's Lagrangian.


\subsection{
Energy in a fixed frame calculated from definition} \label{fixdef}

 Now , instead of  (\ref{kdv1}) we consider the second order KdV equation, that is
 (\ref{etaab})
called by Marchant \& Smyth \cite{MS} "extended KdV".

In section \ref{energy} A,   
total energy of the wave governed by KdV equation, that is the equation (\ref{kdv1}) with terms only up to first order in small parameters was obtained in (\ref{EcRz1}).
In calculation according to eq. (\ref{etaab}) the potential energy  is expressed by the same formula (\ref{EpRz1}) as previously for KdV equation.
In the expression for kinetic energy 
the velocity potential has to be expanded to second order in small parameters
\begin{equation}\label{velp}
\phi= f -\frac{1}{2}\beta y^2 f_{xx}+\frac{1}{24}\beta^2 y^4 f_{xxxx},
\end{equation}
with derivatives
\begin{equation}\label{velpx}
\left\{ \begin{array}{lll} 
\phi_x & = & f_x -\frac{1}{2}\beta y^2 f_{xxx}+\frac{1}{24}\beta^2 y^4 f_{xxxxx}, \\
 & & \\ 
\phi_y & = & -\beta y f_{xx}+\frac{1}{6}\beta^2 y^3 f_{xxxx}. \end{array} \right.
\end{equation}
Integrating over $y$ and retaining terms up to fourth order  yields
\begin{eqnarray}\label{EK4}
T&=&\frac{1}{2} \rho g h^2 l \int^{+\infty}_{-\infty} \alpha^2\left[ f_x^2 + \alpha \eta f_x^2 +\frac{1}{3}\beta\left(f_{xx}^2-f_x f_{xxx}  \right) \right.  \nonumber\\
&&  \hspace{15ex}  
+ \alpha\beta (\eta f_{xx}^2 - \eta f_x f_{xxx}) \\ && \left.
 + \beta^2 \left(\frac{1}{20} f_{xxx}^2 -\frac{1}{15} f_{xx} f_{xxxx} + \frac{1}{60} f_{x} f_{xxxxx}  \right)
\right] dx. \nonumber
\end{eqnarray}
Expression (\ref{EK4}) limited to first line gives kinetic energy for KdV equation,
see (\ref{EkRz1}).

Now, we use the expression for $f_x$ (and its derivatives) up to second order, see e.g.\  \cite[Eq.~(2.7)]{MS}, \cite[Eq.~(17)]{KRI}
\begin{eqnarray} \label{f1x2}
f_x &=&  \eta -\frac{1}{4}\alpha\eta^2 + \frac{1}{3}\beta \eta_{xx}+ \frac{1}{8}\alpha^2 \eta^3  \\ &&
+ \alpha\beta\left(\frac{3}{16}\eta_{x}^2 +\frac{1}{2} \eta\eta_{xx}\right)+\frac{1}{10} \beta^2 \eta_{xxxx}. \nonumber
\end{eqnarray}

Insertion (\ref{f1x2}) and its derivatives into (\ref{EK4}) gives
\begin{eqnarray}\label{EK4a}
T&=&\frac{1}{2} \rho g h^2 l \int^{+\infty}_{-\infty} \alpha^2\left[ \eta^2 +\frac{1}{2}  \alpha \eta^3 +\frac{1}{3}\beta\left(\eta_{x}^2+\eta \eta_{xx}  \right) \right. \nonumber \\
&&  \hspace{3ex}   -\frac{3}{16}  \alpha^2 \eta^4 
+ \alpha\beta \left(\frac{29}{24} \eta \eta_{x}^2 + \frac{3}{4} \eta^2 \eta_{xx}\right) \\ && \left.
+ \beta^2 \left(\frac{1}{20} \eta_{xx}^2 +\frac{7}{45} \eta_{x} \eta_{xxx} + \frac{19}{180} \eta \eta_{xxxx}  \right)
\right] dx. \nonumber
\end{eqnarray}
From properties of solutions at $x\to\pm\infty$ terms with $\beta$ and $\beta^2$  in square bracket vanish and the term with $\alpha\beta$ can be written form. Finally one obtains
\begin{equation} \label{EK4b}
T=\frac{1}{2} \rho g h^2 l\!\! \int^{+\infty}_{-\infty}\!\!\! \alpha^2 \!\!\left[ \eta^2
 +\frac{1}{2}  \alpha \eta^3-\frac{3}{16}  \alpha^2 \eta^4 - \frac{7}{24} \alpha\beta \eta \eta_{x}^2   \right]\! dx.
\end{equation}
Then total energy is the sum of (\ref{EpRz1a}) and (\ref{EK4b})
\begin{eqnarray}\label{EcRz2}
E_\mathrm{tot} &=& \rho g h^2 l    \int_{-\infty }^{\infty } \left[\alpha \eta + (\alpha \eta)^2 + \frac{1}{4}(\alpha \eta)^3 \right. \hspace{2ex}\\ && \left.
\hspace{10ex} - \frac{3}{32}(\alpha\eta)^4 - 
\frac{7}{48}\alpha^3\beta \eta \eta_x^2 \right]  dx . \nonumber
\end{eqnarray}

The first three terms are identical as in KdV energy formula (\ref{EcRz1}), the last two terms are new for extended KdV equation (\ref{etaab}).


\subsection{Energy in a fixed frame calculated from Luke's Lagrangian}  \label{fixLuk}

Calculate energy in the same way as in Section \ref{LukL}, C, but  in one order higher. In scaled coordinates Lagrangian density is expressed by 
(\ref{LL}) (here we keep infinite constant term)
\begin{eqnarray}\label{Luke1}
L &=& \rho g h^2 l\left\{ \int^{1+\alpha \eta}_{0} \alpha \left[ \phi_t +\frac{1}{2}\left(\alpha \phi_{x}^2 + \frac{\alpha}{\beta} \phi_y^2 \right) \right]\, dy \right. \nonumber \\ && \left. \hspace{8ex}
+ \frac{1}{2}(1+\alpha\eta)^2 \right\} .
\end{eqnarray}
From  (\ref{velp}) we have 
\begin{equation}\label{phit}
\phi_t = f_t-\frac{1}{2} \beta y^2 f_{xxt} + \frac{1}{24} \beta^2 y^4 f_{xxxxt}.
\end{equation}
Inserting  (\ref{phit}) and (\ref{velpx}) into (\ref{Luke1}), integrating over $y$ and retaining terms up to third order one obtains (constant term $\frac{1}{2}$ id dropped)
\begin{eqnarray} \label{Lukey}
\frac{L}{ \rho g h^2 l } &=&  \alpha\left\{ (\eta+f_t)  + \alpha \left(\frac{1}{2}\eta^2 +\eta f_t + \frac{1}{2} f_x^2 \right) - \frac{1}{2}\beta f_{xxt} \right. 
\nonumber \\ &&
 + \frac{1}{2}\alpha^2 \eta f_x^2
+\alpha\beta\left(\frac{1}{6}f_{xx}^2 -\frac{1}{2}\eta f_{xxt}-\frac{1}{6}f_{x} f_{xxx} \right) \nonumber  \\ &&
 + \frac{1}{120} \beta^2 f_{xxxxt}
 +\! \frac{1}{2} \alpha^2\beta\left( \eta f_{xx}^2 \!-\! \eta^2 f_{xxt} \!-\! \eta f_{x} f_{xxx} \right)\nonumber  \\ &&
+  \alpha\beta^2\left( \frac{1}{40} f_{xxx}^2 - \frac{1}{30}f_{xx} f_{xxxx} \right.\\ &&\left.  \left. + \frac{1}{24} \eta f_{xxxxt} +\frac{1}{120}f_{x} f_{xxxxx}\right) -\beta^3
\frac{f_{xxxxxxt}}{5040} \right\}
.\nonumber
\end{eqnarray}
The the Hamiltonian density 
$$ H = f_t\frac{\partial L}{\partial f_t} + f_{xxt}\frac{\partial L}{\partial f_{xxt}}+ f_{(4x)t}\frac{\partial L}{\partial f_{(4x)t}} +  f_{(6x)t}\frac{\partial L}{\partial f_{(6x)t}} -L$$
is
\begin{eqnarray} \label{HLyy}
\frac{H}{ \rho g h^2 l } &=&-\alpha\eta -\frac{1}{2}\alpha^2 \left(\eta^2 +f_{x}^2 \right) \nonumber \\ &&
-\frac{1}{2}\alpha^3 \eta f_{x}^2 + \alpha^2\beta  \left(-\frac{1}{6}f_{xx}^2 + \frac{1}{6}f_{x} f_{xxx}\right) \nonumber \\ &&
  + \alpha^3\beta  \left(-\frac{1}{2} \eta f_{xx}^2 + \frac{1}{2} \eta f_{x} f_{xxx}\right) \\ &&
 + \alpha^2\beta^2  \left( -\frac{1}{40} f_{xxx}^2+ \frac{1}{30}f_{xx} f_{xxxx} -\frac{1}{120}f_{x} f_{xxxxx}
\right). \nonumber
\end{eqnarray}
Now, we use $f_x$in the second order  (\ref{f1x2}) and its derivatives. Insertion these expressions into  (\ref{HLyy}) nd retention terms up to thired order yields
\begin{eqnarray} \label{HLyy1}
\frac{H}{ \rho g h^2 l } &=& - \alpha\eta - \alpha^2 \eta^2-\frac{1}{4}\alpha^3\eta^3+ \frac{3}{32} \alpha^4\eta^4\nonumber \\ &&
 + \alpha^2\beta  \left( -\frac{1}{6}\eta_{x}^2 -\frac{1}{6}\eta \eta_{xx} \right) 
 \\ && 
+ \alpha^3\beta  \left(-\frac{29}{48} \eta \eta_{x}^2 - \frac{3}{8}\eta^2 \eta_{xx}\right)\nonumber \\ && 
 + \alpha^2\beta^2  \left(  -\frac{1}{40} \eta_{xx}^2- \frac{7}{90}\eta_{x} \eta_{xxx} -\frac{19}{360}\eta \eta_{xxxx}
\right).\nonumber
\end{eqnarray}
The energy is obtained by integration of (\ref{HLyy1}) over  $x$ (using integration by parts and properties of  $\eta$ and its derivatives at $x\to\pm\infty$). Then terms with $\alpha\beta$ and $\alpha\beta^2$ vanish. The final result is
\begin{eqnarray} \label{ELukNieRuch}
E &=& - \rho g h^2 l \int_{-\infty}^{+\infty}\left[ \alpha\eta +(\alpha\eta)^2 +\frac{1}{4}(\alpha\eta)^3 \right. \\ && \left. \hspace{13ex}
- \frac{3}{32}(\alpha\eta)^4 -\frac{7}{48}\alpha^3\beta\eta\eta_x^2 \right] dx, \nonumber 
\end{eqnarray}
the same as (\ref{EcRz2}) but with the opposite sign.


\subsection{Energy in a moving frame from definition} \label{movdef}

Let us follow arguments given by Ali and Kalisch \cite[Sec. 3]{Kalisch} and used already in Section \ref{energy} B. Working in a moving frame one has to replace $\phi_x$ by the horizontal velocity in a moving frame, that is, $\phi_x-\frac{1}{\alpha}$. Then in a frame moving with the sound velocity we have 
\begin{equation}\label{velpxm}
\left\{ \begin{array}{lll}
\phi_x & = & f_x -\frac{1}{2}\beta y^2 f_{xxx}+\frac{1}{24}\beta^2 y^4 f_{xxxxx}- \frac{1}{\alpha},  \\ & & \\
\phi_y & = &  -\beta y f_{xx}+\frac{1}{6}\beta^2 y^3 f_{xxxx}. \end{array} \right.
\end{equation}
Then the expression under integral over $y$ in (\ref{EnKinNZm}) becomes (in the following terms up to fourth order are kept)
\begin{eqnarray}\label{podcal}
 \left( \alpha^2 \phi_{x}^2  \right. &+& \left. \frac{\alpha^2}{\beta} \phi_y^2 \right)  = 1 - 2 \alpha f_x + \alpha^2 f_x^2 +y^2 \alpha^2 \beta f_{xx}^2  \\
&+& y^2 \alpha\beta f_{xxx} - y^2 \alpha^2\beta f_x f_{xxx}
- \frac{1}{12} y^4 \alpha\beta^2  f_{xxxxx} \nonumber \\
&+&  y^4 \alpha^2\beta^2  \left(\frac{1}{4} f_{xxx}^2 -\frac{1}{3}f_{xx} f_{xxxx} + \frac{1}{12}f_{x} f_{xxxxx} \right) .\nonumber
\end{eqnarray}

After integration over $y$  one obtains
\begin{eqnarray}\label{Ekfx}
T & = & \frac{1}{2} \rho g h^2 l \int^{+\infty}_{-\infty} \left[ 1 + \alpha
\left(\eta - 2 f_x \right) +\alpha^2 \left(-2\eta f_x+f_x^2 \right) \right. \nonumber \\  &   &    \hspace{8ex}
+ \frac{1}{3} \alpha\beta f_{xxx} + \alpha^3 \eta f_x^2  - \frac{1}{60}\alpha\beta^2 f_{xxxxx}  \nonumber \\ &   &  \hspace{8ex}
 +\alpha^2\beta\left(\frac{1}{3} f_{xx}^2 + \eta f_{xxx}  - \frac{1}{3} f_{x}  f_{xxx}\right) \nonumber  \\  &   &  \hspace{8ex}
+\alpha^3\beta\left(\eta f_{xx}^2 + \eta^2 f_{xxx} - \eta f_{x} f_{xxx}\right) \nonumber \\
 &   &   \hspace{8ex} + \alpha^2\beta^2 \left( \frac{1}{20}f_{xxx}^2 -\frac{1}{15}
f_{xx}f_{xxxx} \right. \\  &   & \hspace{14ex} \left.  \left.
-\frac{1}{12} \eta f_{xxxxx}+\frac{1}{60}f_{x}f_{xxxxx}
\right)  \right] dx. \nonumber
\end{eqnarray}

Then insertion $f_x$ (\ref{f1x2}) and its derivatives yields
\begin{eqnarray}\label{EkMFr}
T &=& \frac{1}{2} \rho g h^2 l \int^{+\infty}_{-\infty} \left[
 - \alpha\eta -\frac{1}{2}\alpha^2\eta^2 -\frac{1}{3}\alpha\beta \eta_{xx}  \right.   \nonumber\\ & + &
\frac{3}{4}\alpha^3\eta^3 - \alpha^2\beta \left(\frac{5}{24}\eta_x^2 + \frac{1}{2}\eta\eta_{xx} \right) 
-\frac{19}{180}\alpha\beta^2 \eta_{xxxx} \nonumber  \\ &-&
\frac{7}{16}\alpha^4\eta^4 + \alpha^3\beta  \left(\frac{7}{12} \eta\eta_x^2+ \frac{3}{8} \eta^2\eta_{xx} \right) \\  &+& 
 \alpha^2\beta^2\left(\frac{11}{30} \eta_{xx}^2 +\frac{233}{360} \eta_x\eta_{xxx} + \frac{119}{360} \eta\eta_{xxxx}  \right)  \nonumber\\ &+& \left.
 \frac{1}{36} \alpha\beta^3 \eta_{xxxxxx} \right] dx,\nonumber
\end{eqnarray}
where constant term is dropped. Using properties of solutions at $x\to\pm\infty$  this expression can be simplified to 
\begin{eqnarray}\label{EkMFr1}
T&=&\frac{1}{2} \rho g h^2 l \!\!\int^{+\infty}_{-\infty}\!\! \left[
 - \alpha\eta -\frac{1}{2}\alpha^2\eta^2  +\frac{3}{4}\alpha^3\eta^3 - \frac{7}{16}\alpha^4\eta^4 \right.\nonumber \\ && \left.
+ \frac{7}{24}\alpha^2\beta\, \eta_x^2 + \frac{1}{12} \alpha^3\beta \,  \eta\eta_x^2 
+ \frac{1}{20} \alpha^2\beta^2 \eta_{xx}^2 \right]\! dx.
\end{eqnarray}
Then total energy is
\begin{eqnarray} \label{EtMFrRz2}
E_{\mathrm{tot}}&=& \rho g h^2 l \!\!\int^{+\infty}_{-\infty}\!\! \left[
 \frac{1}{2} \alpha\eta +\!\frac{1}{4}\alpha^2\eta^2 +\frac{3}{8}\alpha^3\eta^3- \frac{7}{32}\alpha^4\eta^4   \right.\nonumber \\ &+& \left.
 \!\frac{7}{48}\alpha^2\beta\, \eta_x^2 +\! \frac{1}{24} \alpha^3\beta \,  \eta\eta_x^2 + \!\frac{1}{40} \alpha^2\beta^2 \eta_{xx}^2 \right]\! dx.
\end{eqnarray}

In special case $\alpha=\beta$ this formula simplifies to
\begin{eqnarray} \label{EtMFrab}
E_{\mathrm{tot}} &=& \rho g h^2 l \int^{+\infty}_{-\infty} \left[
 \frac{1}{2} \alpha\eta +\frac{1}{4}\alpha^2\eta^2 +\alpha^3 \left(\frac{3}{8}\eta^3 +\frac{7}{48}\eta_x^2\right) \right.\nonumber \\ && \left.
+ \alpha^4\left(- \frac{7}{32}\eta^4 + \frac{1}{24}\eta\eta_x^2  + \frac{1}{40} \eta_{xx}^2\right) \right] dx.
\end{eqnarray}

\subsection{Energy in a moving frame from Luke's Lagrangian}  \label{movLuk}

Follow considerations in Section \ref{LukL}, but with KdV2 equation (\ref{etaab}).
Transforming into the moving frame through (\ref{ur})
we have now
\begin{equation} \label{ur0}
\phi = f -\frac{1}{2} \beta y^2 f_{\bar{x}\bar{x}}+\frac{1}{24} \beta^2 y^4 f_{\bar{x}\bar{x}\bar{x}\bar{x}}, 
\end{equation}
\begin{equation} \label{ur1b}
\phi_x=f_{\bar{x}} -\frac{1}{2} \beta y^2 f_{\bar{x}\bar{x}\bar{x}}+\frac{1}{24} \beta^2 y^4 f_{\bar{x}\bar{x}\bar{x}\bar{x}\bar{x}}, 
\end{equation}
\begin{equation} \label{ur1a}
 \phi_y = -\beta y f_{\bar{x}\bar{x}}+\frac{1}{6} \beta^2 y^3 f_{\bar{x}\bar{x}\bar{x}\bar{x}},
\end{equation}
\begin{eqnarray} \label{ur2a}
\phi_t &=& -f_{\bar{x}} + \frac{1}{2} \beta y^2 f_{\bar{x}\bar{x}\bar{x}}-\frac{1}{24} \beta^2 y^4 f_{\bar{x}\bar{x}\bar{x}\bar{x}\bar{x}} \\
&& +\alpha (f_{\bar{t}}- \frac{1}{2} \beta y^2 f_{\bar{x}\bar{x}\bar{t}}+\frac{1}{24} \beta^2 y^4 f_{\bar{x}\bar{x}\bar{x}\bar{x}\bar{t}}). \nonumber
\end{eqnarray}

Inserting (\ref{ur0})--(\ref{ur2a}) into (\ref{Luke1}) one obtains Lagrangian density 
in moving frame as (constant term $\frac{1}{2}$ is dropped as previously)
\begin{eqnarray} \label{luk3}
\frac{L}{ \rho g h^2 l } &=& \alpha (\eta- f_{\bar{x}})  + \alpha^2\! \left(\frac{1}{2}\eta^2 + f_{\bar{t}} -\eta f_{\bar{x}} + \frac{1}{2} f_{\bar{x}}^2 \right) \nonumber  \\ &+&
 \frac{1}{6}  \alpha\beta  f_{\bar{x}\bar{x}\bar{x}} 
+  \alpha^3\!\left(\eta f_{\bar{t}} +\frac{1}{2}\eta  f_{\bar{x}}^2\right) -\frac{1}{120}\alpha\beta^2  f_{\bar{x}\bar{x}\bar{x}\bar{x}\bar{x}}
   \nonumber \\ &+&   \alpha^2\beta \!\left(\!\frac{1}{6} f_{\bar{x}\bar{x}}^2 -\frac{1}{6} f_{\bar{x}\bar{x}\bar{t}} \nonumber  
+ \frac{1}{2}\eta f_{\bar{x}\bar{x}\bar{x}} -\frac{1}{6} f_{\bar{x}} f_{\bar{x}\bar{x}\bar{x}}\! \right) \nonumber  \\ &+&
 \alpha^3\beta \! \left(\! \frac{1}{2}\eta f_{\bar{x}\bar{x}}^2-\!\frac{1}{2}\eta f_{\bar{x}\bar{x}\bar{t}}+\!\frac{1}{2} \eta^2  f_{\bar{x}\bar{x}\bar{x}}  -\!\frac{1}{2}\eta f_{\bar{x}} f_{\bar{x}\bar{x}\bar{x}}\!\!\right)    \nonumber \\ &+&
  \alpha^2\beta^2 \left( \frac{1}{40} f_{\bar{x}\bar{x}\bar{x}}^2 - \frac{1}{30} f_{\bar{x}\bar{x}}f_{\bar{x}\bar{x}\bar{x}\bar{x}} + \frac{1}{120}f_{\bar{x}\bar{x}\bar{x}\bar{x}\bar{t}}  \right. \nonumber  \\ &&  \hspace{5ex} \left.
- \frac{1}{24} \eta f_{\bar{x}\bar{x}\bar{x}\bar{x}\bar{x}}  + \frac{1}{120} f_{\bar{x}} f_{\bar{x}\bar{x}\bar{x}\bar{x}\bar{x}}
\right)   .
\end{eqnarray}

Then Hamiltonian density
\begin{equation} \label{ham}
H= f_{\bar{t}} \frac{\partial L}{\partial f_{\bar{t}}} +  f_{\bar{x}\bar{x}\bar{t}} \frac{\partial L}{\partial f_{\bar{x}\bar{x}\bar{t}}} +f_{\bar{x}\bar{x}\bar{x}\bar{x}\bar{t}} \frac{\partial L}{\partial f_{\bar{x}\bar{x}\bar{x}\bar{x}\bar{t}}} -L
\end{equation}
after insertion of (\ref{luk3}) into (\ref{ham}) yields
\begin{eqnarray} \label{ham3}
\frac{H}{ \rho g h^2 l } &=& \alpha\left(-\eta +f_{\bar{x}}\right) + \alpha^2\left(-\frac{1}{2}\eta^2 + \eta f_{\bar{x}} -\frac{1}{2} f_{\bar{x}}^2 \right) \nonumber
\\ &-&
\frac{1}{6} \alpha \beta f_{\bar{x}\bar{x}\bar{x}} - \frac{1}{2} \alpha^3 \eta f_{\bar{x}}^2 + \frac{1}{120} \alpha\beta^2 f_{\bar{x}\bar{x}\bar{x}\bar{x}\bar{x}}
\nonumber  \\ &+&
 \alpha^2\beta \left(-\frac{1}{6} f_{\bar{x}\bar{x}}^2 - \frac{1}{2}\eta  f_{\bar{x}\bar{x}\bar{x}} +\frac{1}{6} f_{\bar{x}} f_{\bar{x}\bar{x}\bar{x}} \right)  \\ &+& 
  \alpha^3\beta \left(-\frac{1}{2}\eta f_{\bar{x}\bar{x}}^2 -\frac{1}{2}\eta^2  f_{\bar{x}\bar{x}\bar{x}} + \frac{1}{2}\eta f_{\bar{x}}  f_{\bar{x}\bar{x}\bar{x}}\right)
\nonumber \\ &+&
  \alpha^2\beta^2 \left(-\frac{1}{40} f_{\bar{x}\bar{x}\bar{x}}^2 + \frac{1}{30}  f_{\bar{x}\bar{x}} f_{\bar{x}\bar{x}\bar{x}\bar{x}}  \right. \nonumber\\ && \left.\hspace{6ex}
+ \frac{1}{24}\eta f_{\bar{x}\bar{x}\bar{x}\bar{x}\bar{x}} - \frac{1}{120} f_{\bar{x}}f_{\bar{x}\bar{x}\bar{x}\bar{x}\bar{x}}\right) . \nonumber
\end{eqnarray}
In order to express (\ref{ham3}) by $\eta$ only we use $f_{\bar{x}}$ in the form 
(\ref{f1x2}) 
and its derivatives.  It gives
\begin{eqnarray} \label{HAM3}
\frac{H}{ \rho g h^2 l }  &=&  -\frac{1}{4}\alpha^2\eta^2 + \frac{1}{6}\alpha\beta \eta_{\bar{x}\bar{x}} -\frac{3}{8}\alpha^3\eta^3  \\ &+&
  \alpha^2\beta \left(\frac{5}{48} \eta_x^2 + \frac{1}{4} \eta\eta_{xx} \right) 
+\frac{19}{360} \alpha\beta^2 \eta_{\bar{x}\bar{x}\bar{x}\bar{x}}\nonumber \\ &+&
  \frac{7}{32} \alpha^4\eta^4 
-\alpha^3\beta \left(\frac{7}{24}\eta\eta_{\bar{x}}^2 +\frac{3}{16}\eta^2\eta_{\bar{x}\bar{x}} \right) \nonumber \\ &-&
 \alpha^2\beta^2 \left(\frac{11}{60} \eta_{\bar{x}\bar{x}}^2 + \frac{233}{720}\eta_{\bar{x}}\eta_{\bar{x}\bar{x}\bar{x}}+ \frac{119}{720}\eta \eta_{\bar{x}\bar{x}\bar{x}\bar{x}} \right)\nonumber \\ &-&
 \frac{1}{72} \alpha \beta^3 \eta_{\bar{x}\bar{x}\bar{x}\bar{x}\bar{x}\bar{x}} \nonumber 
\end{eqnarray}

Then energy is given by the integral
\begin{eqnarray} \label{En3}
E &=& \varrho g h^2 l \!\int_{-\infty}^{+\infty}\! \left[-\frac{1}{4}\alpha^2\eta^2 + \frac{1}{6}\alpha\beta \eta_{\bar{x}\bar{x}} -\frac{3}{8}\alpha^3\eta^3 \right.  \\ &&
 + \alpha^2\beta \left(\frac{5}{48} \eta_x^2 + \frac{1}{4} \eta\eta_{xx} \right) 
+\frac{19}{360} \alpha\beta^2 \eta_{\bar{x}\bar{x}\bar{x}\bar{x}}+ \frac{7}{32} \alpha^4\eta^4  \nonumber \\ &&
-\alpha^3\beta \left(\!\frac{7}{24}\eta\eta_{\bar{x}}^2 +\frac{3}{16}\eta^2\eta_{\bar{x}\bar{x}} \right)   - \frac{1}{72} \alpha \beta^3 \eta_{\bar{x}\bar{x}\bar{x}\bar{x}\bar{x}\bar{x}}  \nonumber \\ &&\left.
 - \alpha^2\beta^2 \left(\frac{11}{60} \eta_{\bar{x}\bar{x}}^2 + \frac{233}{720}\eta_{\bar{x}}\eta_{\bar{x}\bar{x}\bar{x}}+\! \frac{119}{720}\eta \eta_{\bar{x}\bar{x}\bar{x}\bar{x}}\!\! \right) \right] dx . \nonumber 
\end{eqnarray}

From properties of solution integrals of terms with $\alpha\beta, \alpha\beta^2,  \alpha\beta^3$ vanish and terms with $\alpha^2\beta, \alpha^3\beta, \alpha^2\beta^2$ can be simplified.
Finally, energy is given by the follwing expression
\begin{eqnarray} \label{En3b}
E &=& \varrho g h^2 l \int_{-\infty}^{+\infty} \left[-\frac{1}{4}\alpha^2\eta^2-\frac{3}{8}\alpha^3\eta^3 + \frac{7}{32} \alpha^4\eta^4 \right.  \\ && \left. \hspace{4ex}
 -\frac{7}{48}\alpha^2\beta \eta_x^2 -\frac{1}{24}\alpha^3\beta \eta \eta_{\bar{x}}^2 - \frac{1}{40}\alpha^2\beta^2  \eta_{\bar{x}\bar{x}}^2  \right] dx.\nonumber 
\end{eqnarray}

In special case when  $\beta= \alpha$~  the result is 
\begin{eqnarray} \label{EN3}
E &=& \varrho g h^2 l \int_{-\infty}^{+\infty} \left[-\frac{1}{4}\alpha^2\eta^2
-\alpha^3\left( \frac{3}{8}\eta^3 +\frac{7}{48} \eta_x^2 \right) \right. \nonumber \\ && \left.\hspace{4ex} + \alpha^4
\left(\frac{7}{32} \eta^4 - \frac{1}{24}\eta \eta_{\bar{x}}^2 -\frac{1}{40} \eta_{\bar{x}\bar{x}}^2  \right)
  \right] dx  .
\end{eqnarray}

If the invariant term $I^{(1)}\equiv \int \alpha\eta\,dx$ is dropped in (\ref{EcRz2}) or (\ref{ELukNieRuch}) then the energy calculated in the moving frame  (\ref{En3b})  have the same value but with oposite sign.

\begin{figure}[tbh]
\begin{center}
\resizebox{0.999\columnwidth}{!}{\includegraphics{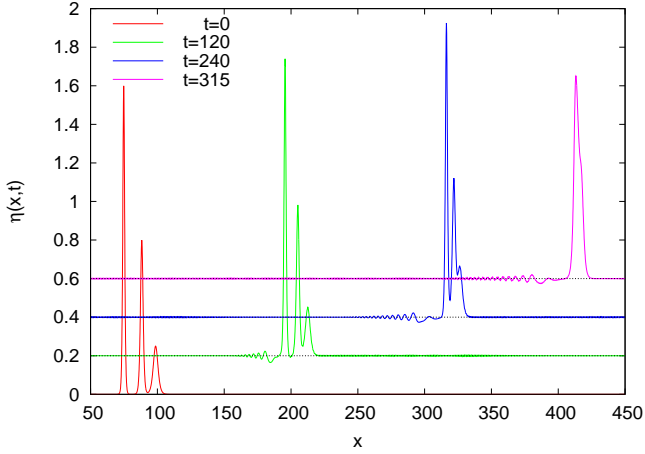}}
\end{center}
\vspace{-5mm}
\caption{Example of time evolution of 3-soliton solution.} 
 \label{3sN}
\end{figure}

\subsection{Numerical tests}\label{NumT}

\subsubsection{Fixed frame}\label{FCS}
\begin{figure}[tbh]
\begin{center}
\resizebox{0.999\columnwidth}{!}{\includegraphics{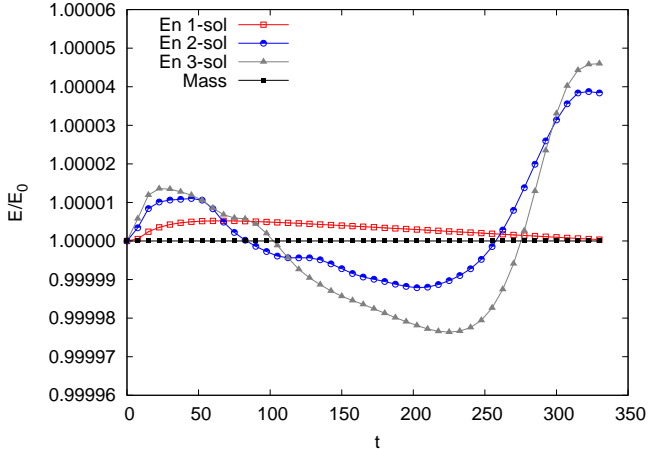}}
\end{center}
\vspace{-5mm}
\caption{Energy (non)conservation for the extended KdV equation in the fixed frame (\ref{etaab}). Symbols represent values of the total energy given by formulas (\ref{EcRz2}) or (\ref{ELukNieRuch}). Full square symbols represent the invariant $I^{(1)}$.} 
 \label{EnonR}
\end{figure}

\begin{figure}[bth]
\begin{center}
\resizebox{0.999\columnwidth}{!}{\includegraphics{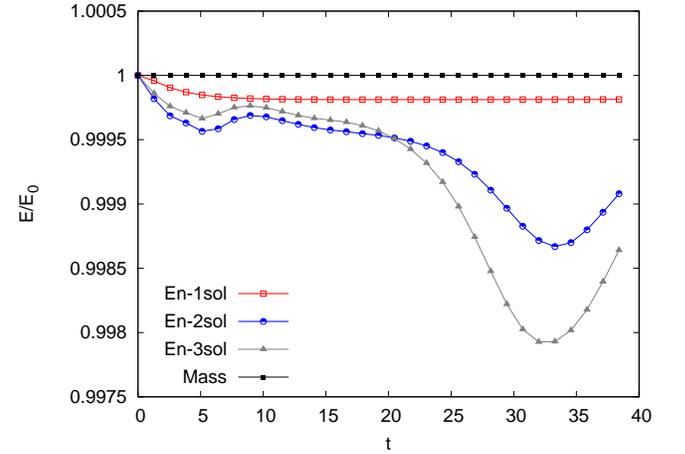}}
\end{center}
\vspace{-5mm}
\caption{Energy (non)conservation for the extended KdV equation in the moving frame (\ref{etaabM}). Symbols represent values of the total energy given by the formula (\ref{EtMFrRz2}). Full square symbols represent the invariant $I^{(1)}$.} 
 \label{EnonN}
\end{figure}

In order to check energy conservation for the extended KdV equation (\ref{etaab}) we performed several numerical tests. 
First, discuss energy conservation in a fixes frame. 
We calculated time evolution governed by the equation (\ref{etaab}) of waves which initial shape was given by 1-, 2- and 3-soliton solutions of the KdV (first order) equations. 
For presentation the following initial conditions were chosen. 3-soliton solution have amplitudes 1.5, 1 and 0.25,  2-soliton solution have amplitudes 1 and 0.5 and 1-soliton solution the amplitude 1. The changes of energy presented in Figs.\ \ref{EnonN} and \ref{EnonR} are qualitatively the same also for different amplitudes.
An example of such time evolution for 3-soliton solution is presented in Fig. \ref{3sN}.
  
Time range in Fig.\ \ref{3sN} contains initial  shape of 3-soliton solution with almost separated solitons at $t=0$, intermediate shapes and almost ideal overlap of solitons at $t=315$. In order to do not obscure details the subsequent shapes are shifted verticaly with respect to the previous ones.
 Note additional slower waves after the main one which are generated by second order terms of the equation (\ref{etaab}), that were already discussed in \cite{KRI}.

We see that the total energy for waves which move according to the extended KdV equation is {\bf not} conserved. Although energy variations are generally small (in time range considered they do not extend 0.001\%, 0.004\% and 0.005\% for 1-, 2-, 3-soliton waves, respectively) they increase with more complicated waves. For additional check of numerics the invariant $
I^{(1)}=\int_{-\infty}^{+\infty} \alpha\eta(x,t) dx$ for the eaquation (\ref{etaab}) was plotted as {\sf Mass}. In spite of approximate integration the value of $I^{(1)}$ was obtained constant up to 10 digits for all initial conditions.

\subsection{Moving frame} \label{MCS}

Here we present variations of the energy calculated in a moving frame. The time evolution of the wave is given by the equation (\ref{etaab}) transformed with (\ref{ur}), that is the equation
\begin{eqnarray} \label{etaabM}
\eta_{\bar{t}} \!&\!+\!& \! \frac{3}{2}\eta\eta_{\bar{x}} + \frac{1}{6}\frac{\beta}{\alpha} \eta_{3\bar{x}} \\ &-&
\frac{3}{8}\alpha\,\eta^2\eta_{\bar{x}} +\beta\,\left(\frac{23}{24}\eta_{\bar{x}}\eta_{2{\bar{x}}}+\frac{5}{12}\eta\eta_{3\bar{x}} \right)+\frac{19}{360}\frac{\beta^2}{\alpha}\eta_{5\bar{x}} =0.  \nonumber
\end{eqnarray}

The time range of the evolution was chosen for a convenient comparison with the numerical results obtained in fixed reference frame, that is 2- and 3-soliton waves move from separate solitons to fully colliding time instant. The convention of symbols is the same as in Fig. \ref{EnonR}. the energy is calculated according to the formula (\ref{EtMFrRz2}).
In moving coordinate system energy variations are even greater than in the fixed reference frame, because in the time period considered it approaches values of 0.02\%,  0.12\% and 0.2\% for 1-, 2- and 3-soliton waves, respectively. This increase of relative time variations of energy can not be atributed only to two times smaller leading term ($\frac{1}{2}\alpha\eta$) in (\ref{EtMFrRz2}) with respect to  (\ref{EcRz2}). Again, in spite of approximate integration the value of $I^{(1)}$ was obtained constant up to 10 digits for all initial conditions.

\subsubsection{Conclusions for extended KdV equation}

We calculated energy of the fluid governed by the extended KdV equation (\ref{etaab}) in two cases:
\begin{enumerate}
\item In a fixed frame (sections \ref{fixdef} and \ref{fixLuk}).
\item In the frame moving with the sound velocity (sections \ref{movdef} and \ref{movLuk}).
\end{enumerate}
In both cases we calculated energy using two methods: from definition and from Luke's Lagrangian. Both methods give consistent results. For fixed frame energies (\ref{EcRz2}) and  (\ref{ELukNieRuch}) are the same.
For moving frame the energy calculated from the definition contains one term more then energy calculated from Luke's Lagrangian, but this term ($\int \alpha\eta\, dx$)  is the invariant $I^{(1)}$. When this term is dropped both energies in moving coordinate system (\ref{EtMFrRz2}) and (\ref{En3b}) are the same and energies in both coordinate systems differ only by sign.

The general conclusion concerning energy conservation for shallow water wave problem can be formulated as follows. Since there exists the Lagrangian of the system (Luke's Lagrangian) then exact solutions of Euler equations have to conserve energy. However, when approximate equations of different orders resulting from exact Euler equations are considered, energy conservation is not a priori determined. The KdV equations obtained in  first order approximation has a miraculous property, an infinite number of invariants with energy among them. However, this astonishing property is lost in second order approximation to Euler equations and energy in this order may be conserved only approximately.

\appendix

\section{
} \label{app1}

The simplest,  {\sl mathematical}~ form of the KdV equation is obtained from (\ref{kdv1})  by passing to the moving frame with additional scaling 
\begin{equation} \label{tr}
\bar{x} =\sqrt{\frac{3}{2}}(x-t), \qquad \bar{t}=\frac{1}{4}\sqrt{\frac{3}{2}}\,\alpha\,t, \qquad u = \eta,
\end{equation}
which gives a standard, {\sl mathematical}~ form of the KdV equation
\begin{eqnarray}\label{kdvm}
u_{\bar{t}} + 6\,u\,u_{\bar{x}} + \frac{\beta}{\alpha}u_{\bar{x}\bar{x}\bar{x}} &=& 0,
\quad \mbox{or} \nonumber \\ 
u_{\bar{t}} + 6\,u\,u_{\bar{x}} + u_{\bar{x}\bar{x}\bar{x}} &=& 0 \quad \mbox{for} \quad  \beta=\alpha.
\end{eqnarray}
Equations (\ref{kdvm}), particularly with ~$\beta=\alpha$~ are favored by mathematicians,  see, e.g.\ \cite{Lax}. This form of KdV is the most convenient for ISM (the Inverse Scattering Method, see, e.g.\ \cite{GGKM,AS,AC}).

For the moving reference frame, in which the KdV equation has a standard (mathematical) form (\ref{kdvm}),  the invariant ~$I^{(3)}$~ is slightly different. To see this difference  denote the lhs of  (\ref{kdvm}) by  ~$\mbox{KDVm}(x,t)$ and construct
$$3 \eta^2 \times \mbox{KDVm}(x,t) - \frac{\beta}{\alpha} \eta_x \times \frac{\partial} {\partial x} \mbox{KDVm}(x,t)=0.$$ 
Then after simplifications one obtains
\begin{eqnarray} \label{i3r}
\frac{\partial} {\partial t}\left( \eta^3 -\frac{1}{2}  \frac{\beta}{\alpha}  \eta_x^2  \right) + \frac{\partial} {\partial x}\left[\frac{9}{2} \eta^4 -6 \frac{\beta}{\alpha}\eta\eta_x^2  \right. \hspace{6ex} && \\ \left. +3 \frac{\beta}{\alpha}\eta^2\eta_{xx}
- \frac{1}{2} \left(\frac{\beta}{\alpha} \eta_{xx}\right)^2+\left(\frac{\beta}{\alpha}\right)^2 \eta_{x}\eta_{xxx} \right] &=& 0, \nonumber 
\end{eqnarray}
which implies the invariant ~$I^{(3)}$ in the following form
\begin{eqnarray} \label{i3ruch}
I^{(3)}_{\mathrm{moving~frame}}\!&\!=\!&\! \int_{-\infty}^{\infty}\! \left( \eta^3 -\frac{1}{2}  \frac{\beta}{\alpha} \eta_x^2 \right) dx = \mbox{const} \quad \mbox{or} \\
I^{(3)}_{\mathrm{moving~frame}}\!&\!=\!&\!\int_{-\infty}^{\infty} \!\left( \eta^3 -\frac{1}{2}  \eta_x^2 \right) dx = \mbox{const}
\quad \mbox{for} \quad \beta=\alpha . \nonumber
\end{eqnarray}
We see, however, that the difference between (\ref{i3ruch}) and (\ref{i3Em}) is caused by additional scaling. 

In the Lagrangian approach as described in Sect.~\ref{varia}, the substitution ~$u=\varphi_x$ into (\ref{kdvm}) gives
\begin{equation} \label{e1pa}
\varphi_{xt} +6\, \varphi_x \varphi_{xx}+\varphi_{xxxx}=0\,.
\end{equation}
Then the appropriate Lagrangian density for equation (\ref{kdvm}) with ($\alpha=\beta$) is
\begin{equation} \label{L1W}
\mathcal{L}_\mathrm{standard~KdV} := -\frac{1}{2} \varphi_t\varphi_x-\varphi_x^3+\frac{1}{2} \varphi_{xx}^2 \, .
\end{equation}
Indeed, the Euler--Lagrange equation obtained from the Lagrangian (\ref{L1W}) is just (\ref{e1pa}).

The Hamiltonian for KdV  (\ref{kdvm}) can be found e.g.\ in \cite{PL_A173}. Defining generalized momentum ~$\displaystyle\pi=\frac{\partial  \mathcal{L}}{\partial \varphi_{t}}$, where ~$\mathcal{L}$ is given by (\ref{L1W}),
one obtains
\begin{eqnarray} \label{Hh1a}
H &=& \int_{-\infty}^{\infty} \left[\pi \dot{\varphi}-\mathcal{L}\right] dx=  \int_{-\infty}^{\infty} \left[\frac{\partial  \mathcal{L}}{\partial \varphi_{t}}  \varphi_t-\mathcal{L}\right] dx  \\
&=& \int_{-\infty}^{\infty} \left[\varphi_{x}^3-\frac{1}{2} \varphi_{xx}^2\right] dx = \int_{-\infty}^{\infty} \left[\eta^3-\frac{1}{2} \eta_{x}^2\right] dx \,.\nonumber
\end{eqnarray}
This is the same invariant as ~$I^{(3)}_\mathrm{moving~frame}$ in (\ref{i3ruch}).

\section{
} \label{app2}

The set of Euler equations for irrotational motion of an incompresible and inviscid fluid can be written (neglecting surface tension) in dimensionless form:
\begin{eqnarray} \label{eu1}
\nabla^2\phi &=& 0 \\ \label{eu2}
\phi_z  &=& 0 \quad \mbox{on} \quad z=0  \\ \label{eu3}
\eta_t+\phi_x\eta_x- \phi_x  &=& 0 \quad \mbox{on} \quad z=1+\eta  \\ \label{eu4}
\phi_t +\frac{1}{2}\left(\phi_x^2+\phi_z^2 \right)+\eta  &=&0 \quad \mbox{on} \quad z=1+\eta.
\end{eqnarray}

We look for solutions to the Laplace equation (\ref{eu1}) in the form
\begin{equation} \label{ffi}
\phi = \sum_{n=0}^{\infty} z^n\,f^{(n)}(x,y,t)
\end{equation}
yielding
\begin{equation} \label{ffi1}
\sum_{n=0}^{\infty} \left[n(n-1)z^{n-2}\,f^{(n)}+ z^n \nabla^2f^{(n)}\right]=0.
\end{equation}
In two dimensions $(x,z)$ we obtain 
\begin{equation} \label{ffi2}
f^{(n+2)}=\frac{-1}{(n+1)(n+2))}\frac{\partial^2 f^{(n)}}{\partial x^2}.
\end{equation}
The boundary condition at the bottom, $\phi_z=0$ at $z\!=\!0$ implies 
$f^{(1)}=0$ and then all odd $f^{(2k+1)}=0$. Now
\begin{equation} \label{ffi3}
\phi = \sum_{n=0}^{\infty} (-1)^m \frac{z^{2m}}{(2m)!} \frac{\partial^{2m} f}{\partial x^{2m}},
\end{equation}
where $f:=f^{(0)}$. In the stretched coordinates $\partial_x^2=\varepsilon\partial_\xi^2$ so
\begin{equation} \label{ffi4}
\phi =\varepsilon^{\frac{1}{2}} \left(f+  
\sum_{n=0}^{\infty} (-1)^m \frac{z^{2m}}{(2m)!} \left(\varepsilon\partial_\xi^2 \right)^{2m}\,f \right).
\end{equation}
Now both (\ref{eu1}) and (\ref{eu2}) are satisfied. We must also satisfy the boundary conditions on $z=1+\eta$.

In the derivation of KdV and Kadomtsev-Petiashvili \cite{KP} from the Euler equations (\ref{eu1})--(\ref{eu4}) Infeld and Rowlands \cite{EIGR} applied scaling assuming the following relations 
$$\mbox{vawelength~:~depth~:~amplitude~~~as}\quad \varepsilon^{-1/2}:1:\varepsilon.$$ 
They then applied a transformation to a frame moving with velocity of sound. The  coordinates scales as
\begin{equation}\label{xtscal} 
\xi=\varepsilon^{\frac{1}{2}}\,(x-t), \qquad \tau = \varepsilon^{\frac{3}{2}}\,t
 \end{equation}
\begin{equation}\label{xtder}
\partial_t =-\varepsilon^{\frac{1}{2}}\,\partial_\xi+\varepsilon^{\frac{3}{2}}\,\partial_\tau, \qquad \partial_x = \varepsilon^{\frac{1}{2}}\partial_\xi.
\end{equation}
For the wave amplitude and velocity potential the appropriate scaling was
\begin{equation} \label{eteps}
\eta =  \varepsilon\eta^{(1)} + \varepsilon^2 \eta^{(2)} + \ldots ,
\end{equation}
and
\begin{equation} \label{fieps}
\phi =  \varepsilon^{\frac{1}{2}}\phi^{(1)} + \varepsilon^{\frac{3}{2}} \phi^{(2)} + \ldots .
\end{equation}
Then the lowest order expression for $\phi$ is 
\begin{equation} \label{firz1}
\phi \approx \varepsilon^{\frac{1}{2}} f- \varepsilon^{\frac{3}{2}}\frac{z^2}{2}\, f_{\xi\xi}
\end{equation}
Next, Infeld and Rowlands show that in order to simultaneously satisfy (\ref{eu3}) and (\ref{eu4}) the next order contributions to $\eta$ and $\phi$ cancel. It  is enough to keep
\begin{equation} \label{etfi1}
1+\eta=1+\varepsilon\eta^{(1)} \quad \mbox{and} \quad \phi=\varepsilon^{\frac{1}{2}}\phi^{(1)}
\end{equation}
and drop upper index $^{(1)}$ in what follows.


\begin{thebibliography}{99}

\bibitem{KRR}  A.~Karczewska, P.~Rozmej and Ł.~Rutkowski, {\it A new nonlinear equation in the shallow water wave problem},
Physica Scripta {\bf 89}, 054026 (2014). 

\bibitem{KRI}  A.~Karczewska, P.~Rozmej and E.~Infeld, {\it Shallow-water soliton dynamics beyond the Korteweg–de Vries equation},
Phys.\ Rev.\ E {\bf 90}, 012907 (2014).

\bibitem{MS} T.R.~Marchant and N.F.~Smyth, {\it The extended Korteweg--de Vries equation and the resonant flow of a fluid over topography}, 
J.\ Fluid Mech.\  {\bf 221}, 263-288 (1990).

\bibitem{BS} G.I.\ Burde, A.\ Sergyeyev, {\it Ordering of two small parameters in the shallow water wave problem}, 
J.\ Phys.\ A:\ Math.\ Theor. {\bf 46}, 075501 (2013).

\bibitem{Miura} R.M.~Miura, {\it KdV equation and generalizations I. A remarkable explicit nonlinear transformation}, 
J.\ Math. Phys.\ {\bf 9}, 1202-1204 (1968).

\bibitem{MGK} R.M.~Miura, C.S.~Gardner and M.D.~Kruskal,  {\it KdV equation and generalizations II. Existence of conservation laws and constants of motion}, 
J.\ Math. Phys.\ {\bf 9}, 1204-1209 (1968).


\bibitem{DrJ} P.G.\ Drazin and R.S.\   Johnson,  Solitons: An Introduction,  Cambridge University Press, Cambridge, 1989. 

\bibitem{Newell85} A.C.~Newell, {\it Solitons in Mathematics and Physics}, Philadelphia: Society for Industraial and Applied Mathematics, 1985.  

\bibitem{B-S87} J.G.B.~Byatt-Smith, {\it On the change of amplitude of interacting solitary waves}, J.\  Fluid Mech.\   {\bf 182},  495-497 (1987).

\bibitem{SK_PO} S.\ Kichenassamy and P.\ Olver,  {\it Existence and nonexistence of solitary wave solutions to higher-order model evolution equations},
SIAM J.\ Math.\ Anal., {\bf 23}, 1141-1166 (1992). 

\bibitem{MS96}  T.R.~Marchant and N.F.~Smyth, {\it Soliton Interaction for the Korteweg-de Vries equation}, IMA J.\ Appl.\ Math.\   {\bf 56}, 157-176 (1996).

\bibitem{Mar99}  T.R.~Marchant, {\it Coupled Korteweg - de Vries equations describing, to higher-order, resonant flow of a fluid over topography},  Phys.\  Fluids   {\bf 11}, No.\ 7, 1797-1804 (1999).

\bibitem{Mar02}  T.R.~Marchant, {\it High-order interaction of solitary waves on shallow water},  Studies in Appl.\ Math.\   {\bf 109}, 1-17 (2002).

\bibitem{Mar02a}  T.R.~Marchant, {\it Asymptotic solitons for a higher-order modified Korteveg-de Vries Equations},  Phys.\ Rev.\ E   {\bf 66}, 046623(1-8) (2002).

\bibitem{ZouSu} Q.~Zou and CH-H.~Su, {\it Overtaking collision between two solitary waves},  Phys.\  Fluids   {\bf 29}, No.\ 7, 2113-2123 (1986).

\bibitem{TMAB}  E.~Tzirtzilakis, V.\ Marinakis, C.\ Apokis and T. Bountis, {\it Soliton-like solutions of higher order wave equations of Korteweg-de Vries type},  J.\ Math.\ Phys.\    {\bf 43}, No.\ 12, 6151-6165 (2002).

\bibitem{Burde}  G.I.\ Burde, {\it Solitary wave solutions of the higher-order KdV models for bi-directional water waves}, 
Commun.\ Nonlinear Sci.\ Numerical Simulat.\ {\bf 16}, 1314-1328 (2011). 

\bibitem{Hirota72} R.~Hirota, {\it Exact solution of the Korteweg-de Vries equation for multiple collisions of solitons}, Phys.\ Rev.\ Lett.\ {\bf 27}, 1192-1194 (1972).

\bibitem{Hir} R.\ Hirota, {\em  The Direct Method in Soliton Theory}, Cambridge University Press, Cambridge, (2004), first published in Japanese (1992).

\bibitem{Kalisch} A.\ Ali and H.\ Kalisch, {\it On the formulation of mass, momentum and energy conservation in the KdV equation}, 
Acta Appl. Math. {\bf 133}, 113-131 (2014).

\bibitem{Whit} G.B.~Whitham, {\it Linear and Nonlinear Waves}, Wiley, New York, 1974.

\bibitem{PL_A173} F.~Cooper, C.~Lucheroni, H.~Shepard and P.~Sodano, {\it Variational Method for Studying Solitons in the Korteweg-deVries Equation}, 
Phys.\ Lett.\ A {\bf 173} 33-36 (1993). 

\bibitem{Luke} J.C.~Luke, {\it A variational principle for a fluid with a free surface}, 
J.\ Fluid Mech.\ (1967), {\bf 27}, part 2, 395-397.

\bibitem{EIGR} E. Infeld and G. Rowlands, {\it Nonlinear Waves, Solitons and Chaos}, 2nd edition, Cambridge University Press, Cambridge, 2000.

\bibitem{Lax} P.~Lax, {\it Integrals of nonlinear equations of evolution and solitary waves}, 
Comm. Pure Applied Math. {\bf 21} (5), 467–490 (1968).

\bibitem{GGKM} C.S.~Gardner, J.M.~Greene, M.D.~Kruskal, and R.M.~Miura, {\it Method for Solving the Korteweg-deVries Equation}, 
Phys.\ Rev.\ Lett.\ {\bf 19}, 1095-1097 (1967).

\bibitem{AS} M.~Ablowitz, H.~Segur, {\it Solitons and the Inverse Scattering Transform}, SIAM, Philadelphia, 1981.

\bibitem{AC} M.~Ablowitz, P.~Clarkson, {\it Solitons, Nonlinear Evolution Equations and Inverse Scattering}, Cambridge University Press, Cambridge, 1991.

\bibitem{KP} B.B.~Kadomtsev and V.I.~Petviashvili, {\it On the stability of solitary waves in weakly dispersive media},
 Dokl.\ Akad.\ Nauk SSSR {\bf 192}, 753-756 (1970); Sov.\ Phys.\ Dok.\ {\bf 15}, 539-541  (1970).

\bibitem{IRH} E.~Infeld, G.~Rowlands and M.~Hen, {\it Three dimenional stability of KdV waves and  solitons}, 
Acta Phys.\ Polon.\ A {\bf 54}, 123-143, (1978). 

\bibitem{KP_woda} E.~Infeld and G.~Rowlands, {\it  Three-dimensional stability of Korteweg de Vries waves and solitons, II}, 
Acta Phys.\ Polon.\ A {\bf 56}, 329-332, (1979). 
 
\bibitem{KP_plazma} E.~Infeld, {\it Three-dimensional stability of Korteweg de Vries waves and solitons. III. Lagrangian methods, KdV with positive dispersion},
Acta Phys.\ Polon.\ A {\bf 60} 623-643, (1981). 

\bibitem{Benj} T.B.~Benjamin, {\it The stability of solitary waves},
Proc.\ R.\ Soc.\ Lond.\ A  {\bf 328}, 153-183 (1972).

\bibitem{L-H&F} M.S.~Longuet-Higgins and J.D.~Fenton, 
{\it On the mass, momentum, energy and circulation of a solitary wave. II},
Proc.\ R.\ Soc.\ Lond.\ A  {\bf 340}, 471-493, (1974).

\bibitem{InRo} E.~Infeld and G.~Rowlands, 
{\it Stability of nonlinear ion sound waves and solitons in plasma},  
Proc.\ R.\ Soc.\ Lond.\ A  {\bf 366}, 537-554 (1979).




\end{thebibliography}
\end{document}